\documentclass[runningheads]{llncs}

\usepackage[T1]{fontenc}
\usepackage[table]{xcolor}
\usepackage{graphicx}
\usepackage{svg}
\usepackage[normalem]{ulem}
\useunder{\uline}{\ul}{}
\usepackage{blindtext}
\usepackage{multicol}
\usepackage{float}
\usepackage{verbatim}
\usepackage{array}
\usepackage{multirow}
\usepackage{pifont}
\usepackage{amssymb}
\usepackage{mathtext}
\usepackage{booktabs,siunitx} 

\usepackage{placeins}

\usepackage{etoolbox}           
\renewcommand{\bfseries}{\fontseries{b}\selectfont} 
\robustify\bfseries             
\newrobustcmd{\B}{\bfseries}    

\begin{document}
\title{MALITE: Lightweight Malware Detection and Classification for Constrained Devices}
\author{Sidharth Anand\inst{1} \and Barsha Mitra\inst{1} \and Soumyadeep Dey\inst{2} \and  Abhinav Rao\inst{3*}\thanks{\scriptsize{* The authors contributed while being affiliated with BITS Pilani, Hyderabad Campus}} \and Rupsa Dhar\inst{4*} \and Jaideep Vaidya\inst{5}}
\renewcommand\footnotemark{}
\renewcommand\footnoterule{}
\authorrunning{S. Anand et al.}
\titlerunning{MALITE: Lightweight Malware Detection and Classification}
%
\institute{Department of CSIS, BITS Pilani, Hyderabad Campus, Hyderabad, India \\
\email{\{f20191203,barsha.mitra\}@hyderabad.bits-pilani.ac.in}\\
\and Microsoft India Development Center, India \\
\email{Soumyadeep.Dey@microsoft.com}\\
\and Microsoft India (R \& D)  \\
\email{t-raoabhinav@microsoft.com}\\
\and Apple, India  \\
\email{rupsa\_dhar@apple.com}\\
\and MSIS Department, Rutgers University, New Brunswick, NJ, USA \\
\email{jsvaidya@business.rutgers.edu}
}
\maketitle              
\begin{abstract}
Today, malware is one of the primary cyberthreats to organizations. Malware has pervaded almost every type of computing device including the ones having limited memory, battery and computation power such as mobile phones, tablets and embedded devices like Internet-of-Things (IoT) devices. Consequently, the privacy and security of the malware infected systems and devices have been heavily jeopardized. In recent years, researchers have leveraged machine learning based strategies for malware detection and classification. Malware analysis approaches can only be employed in resource constrained environments if the  methods are lightweight in nature. In this paper, we present MALITE, a lightweight malware analysis system, that can classify various malware families and distinguish between benign and malicious binaries. MALITE converts a binary into a gray scale or an RGB image and employs low memory and battery power consuming as well as computationally inexpensive malware analysis strategies. We have designed MALITE-MN, a lightweight neural network based architecture and MALITE-HRF, an ultra lightweight random forest based method that uses histogram features extracted by a sliding window. We evaluate the performance of both on six publicly available datasets (Malimg, Microsoft BIG, Dumpware10, MOTIF, Drebin and CICAndMal2017), and compare them to four state-of-the-art malware classification techniques. The results show that MALITE-MN and MALITE-HRF not only accurately identify and classify malware but also respectively consume several orders of magnitude lower resources (in terms of both memory as well as computation capabilities), making them much more suitable for resource constrained environments.

\end{abstract}
\section{Introduction} 
\label{sec:introduction}
Malicious software or malware is a huge problem worldwide with over 5.5 billion attacks during 2022 \cite{Sonicwall}. Malware is an application that can potentially damage the environment in which it is executed. Cyber criminals propagate and introduce malware into various computing systems mostly using the Internet with the intent of damaging such systems, espionage and information theft thereby violating user security and privacy. Such computing systems include personal desktop computers, laptops, workstations, servers, mobile phones, tablets and even embedded devices like Internet-of-Things (IoT) devices. Out of the aforementioned computing environments, mobile phones, tablets and embedded devices are considered as resource constrained devices with respect to available memory, battery capacity and computational power. Over the past few years, there has been numerous incidents of malware attacks on different types of computing systems, with over 60 million malware attacks against IoT devices \cite{Sonicwall}. 

Therefore, safeguarding these systems against the various malware families is of utmost importance. Researchers have invested a considerable amount of effort in designing strategies for identifying and classifying malware. The rapid development of artificial intelligence in recent years has lead to the growing interest in leveraging machine learning and deep learning models to effectively detect and classify malware \cite{DANGELO202026}, \cite{KABAKUS2022117833}, \cite{KONG2022102514}, \cite{TEKEREK2022102515}. 
Inspite of the emergence of a huge number of methods dedicated towards malware analysis, few of these approaches pay attention to the overhead imposed in terms of memory consumption, battery consumption and computational complexity. These overheads need to be duly accounted for and even optimized if defensive strategies against malware are to be deployed in resource constrained environments like mobile phones, tablets and IoT devices. In fact, in recent years, there has been a growing concern regarding the security of these devices due to the surge in the spread of android and IoT malware. Thus, we need lightweight malware detection and classification techniques to protect such constrained devices.

In this paper, we propose \textbf{M}alware \textbf{A}nalysis using \textbf{Li}gh\textbf{t}weight M\textbf{e}thods or MALITE to distinguish between benign and malware binaries as well as classify the different malware families. MALITE is capable of performing accurate malware analysis and requires much lesser memory and computation power resulting in reduced battery consumption compared to several state-of-the-art methods. The main contributions of the paper are summarized as follows:
\begin{itemize}
\item We propose MALITE to identify malware binaries and categorize the various malware families by transforming the binaries into gray scale or RGB images. 
The underlying lightweight methods employed by MALITE use low cost strategies like histogram computation, random forest classifier and residual bottleneck layers~\cite{mobilenetv2} resulting in two variants, MALITE-HRF and MALITE-MN.
\item We design an ultra lightweight technique MALITE-HRF in terms of both parameter count and computational cost. MALITE-HRF employs a sliding window to extract the histogram feature from an input image. These histogram features are then used by a random forest classifier to categorize malware. To ensure the lightweight nature of the proposed method, we use histogram binning and restrict the number of trees (also called estimators) as well as the height of each estimator in random forest.
\item We also present MALITE-MN, a lightweight neural network architecture designed using computationally inexpensive residual bottleneck layers~\cite{mobilenetv2}. Further, the design of this architecture ensures a low parameter count leading to reduced memory usage.
\item We show the efficacy of MALITE-HRF and MALITE-MN by evaluating them using six open-source datasets like Malimg \cite{malimg}, Microsoft BIG \cite{ronen2018microsoft}, Dumpware10 \cite{dumpware10}, MOTIF \cite{motif}, Drebin \cite{drebin} and CICAndMal2017 \cite{cic_mal_2017}. 
\item We compare our strategies with four existing malware categorization methods that include 3C2D \cite{3c2d}, DTMIC \cite{KUMAR2022103063}, an approach proposed by Wong et al. \cite{9631209} and MalConv2 \cite{malconv} with respect to identification of various malware families and benign vs. malware classification. 
\item The experimental results highlight that our proposed methods are not only accurate for malware analysis but are also extremely lightweight in terms of parameter count, number of multiplication and addition operations performed and model size when compared to the above mentioned four state-of-the-art approaches. 
Specifically, MALITE-MN requires between 226 to 2 times lesser computational overhead while being between 375 to 6 times smaller in size than these existing methods while achieving comparable or even better performance. 
MALITE-HRF is even more lightweight, requiring between 528611 to 5598 times lesser computational overhead while being between 6761 to 107 times smaller in size than these existing methods and still achieves comparable or better performance.  
\end{itemize}

The rest of the paper is organized as follows. Section \ref{sec:related} reviews the existing literature on malware detection and categorization. In Section \ref{sec:method}, we describe MALITE, our lightweight framework for malware analysis. Dataset description and performance evaluation of MALITE are presented in Section \ref{sec:results}. Finally, we conclude the paper in Section \ref{sec:conclusion}.

\section{Related Work} \label{sec:related}

Over the past several decades, researchers have focused on designing techniques for detecting and classifying different types of malware. The strategies employed for performing the detection and classification tasks include static analysis, dynamic analysis, rule-based approach and graph based methods. The rapid progress of research in the field of artificial intelligence, specially machine learning and deep learning, has led to the development of malware analysis and detection strategies using machine learning. An active learning based approach to collect potentially suspicious files in order to update existing malware databases has been proposed in \cite{NISSIM20145843}. Bae et al. \cite{Bae:Summary} propose a machine learning based method to distinguish ransomware from benign files and classify it among other malware families. A deep learning based strategy to classify malware has been proposed in \cite{9455368}. Recently, a rule-based malware detection approach based on the industry standard of YARA rules \footnote{\tiny{https://yara.readthedocs.io/en/stable/}} has been presented in \cite{brengel2021yarix}.

Android is currently one of the most popular operating systems for mobile phones. As a result, android has been extensively targeted for malware injection. Hence, in the recent past, several techniques have been put forth to make android resilient to malware. Li et al. in \cite{Li:2018} have proposed Significant Permission IDentification (SigPID) to identify android malware based on the permissions used by the android applications. The authors in \cite{OU2022102513} have designed a graph based android malware detection framework. Jerbi et al. \cite{JERBI2020101743} have proposed a dynamic malware detection strategy and a genetic algorithm based artificial malware pattern generator. A machine learning based android malware detection approach based on dynamic analysis has been presented in \cite{Mahindru:2021}. A random forest and API call based strategy that can perform benign vs. malicious android app classification has been outlined in \cite{Jung:2018}. Other android malware detection and classification strategies encompassing static analysis, dynamic analysis, API call based features, machine learning and deep learning include \cite{ELAYAN2021847}, \cite{9204665}, \cite{GAO2021102264}, \cite{IADAROLA2021102198}, \cite{KABAKUS2022117833}, \cite{MAPAS}, \cite{8443370}, \cite{KONG2022102514}, \cite{8629067}, \cite{10.1145/3313391}, \cite{PEI2020101792}, \cite{Pekta:2020}, \cite{SASIDHARAN2021101336}, \cite{SURENDRAN2020102483}, \cite{10.1007/978-3-030-02450-5_11}. Researchers have also proposed methods resilient to malware evolution and obfuscation \cite{10.1145/3162625}, \cite{10.1007/978-3-030-80825-9_16}, \cite{10.1145/3372297.3417291}.

A very popular direction of research in malware detection and classification involves converting the malware binaries into gray scale or RGB images and then classifying these images. Baptista et al. \cite{Baptista:2019} have proposed a deep learning based method for identifying malware by transforming the malware binaries into RGB images. The authors in \cite{8887303} use a Convolutional Neural Network (CNN) based architecture to detect malware after converting the malware files into gray scale and color images. Vasan et al. \cite{VASAN2020107138}, \cite{VASAN2020101748} have also developed image based malware classification techniques that use CNN. In \cite{PINHERO2021102247}, the authors have converted the malware and benign files into gray scale and color images and have used CNN to perform malware vs. benign classification.  
An image and deep transfer learning based malware categorization strategy has been presented in \cite{KUMAR2022103063}. The work in \cite{YUAN2020101740} has transformed the malware binaries into images and have used deep convolutional neural network to classify the images. Other image and deep learning based strategies to distinguish between malware and benign files and to determine the various malware families include \cite{CHAGANTI2022103306}, \cite{BYTEcodeimage}, \cite{9185490}, \cite{JIAN2021102400}, \cite{9356071}, \cite{LIU2020101682}, \cite{3c2d}, \cite{malconv}, \cite{10.1145/3442520.3442522}, \cite{TEKEREK2022102515}, \cite{HIT4Mal}, \cite{9631209}, \cite{XIAO2021102420}. 

The image based strategies for malware analysis have also been applied for android malware. In \cite{nver:2020}, Ünver et al. have developed an android malware classification method by converting the android application files into gray scale images. DEXRAY, a CNN-based approach for malware detection that converts the dex files of the android apps into gray scale images has been presented in \cite{10.1007/978-3-030-87839-9_4}. In \cite{8626828}, the authors have developed a method to identify android malware by converting the malware files into gray scale images and using GIST features. \cite{DANGELO202026} outlines an autoencoder based android malware classification technique that produces an image representation of the API call sequences of various android apps. Several other similar image based approaches are \cite{8955840}, \cite{9136685}, \cite{Mercaldo:2020}, \cite{YADAV2022102622}.

In spite of the presence of the wide array of works on malware analysis, few of them take into consideration the memory overhead and computational complexity of the classification models. In fact, very few of the existing approaches attempt to make the detection and classification tasks lightweight so that the resulting models are suitable for mobile and embedded devices. Some of the methods like \cite{10.1145/3162625} and \cite{10.1145/3313391} highlight the lightweight property in terms of the feature vector size and execution time. However, to the best of our knowledge, none of the current malware analysis works actually showcase the lightweight nature of the detection and classification models in terms of the memory consumed, parameter count and the number of operations performed. This paper attempts to bridge this gap by presenting MALITE for resource constrained devices.

\section{Proposed Approach} \label{sec:method}

In this section, we present our lightweight methods suitable for embedded or memory constrained devices to classify malware.
Our proposed methods perform malware identification and categorization by converting the malware binary into an image. 
The image conversion method is described in Sub-section~\ref{sub-sec:BV}. 
The converted malware images are then classified into malware families based on the features extracted from these images. 
We propose an end-to-end malware classification technique by designing a lightweight Convolutional Neural Network (CNN), and another computationally efficient feature extraction based classification method. 
The proposed classification methods are discussed in Sub-section~\ref{sub-sec:MALITE}.

\subsection{Binary Visualization} \label{sub-sec:BV}

A visual representation of the internal static structure of a binary file can be obtained by converting the binary file into an array of 8-bit unsigned integers. This array forms an image-like plot of the binary file fragments and is known as byteplot. 
This technique was originally introduced in~\cite{Conti20081} and is an efficient method to interpret binary files. 
Byteplot was later applied by Nataraj et al.~\cite{malimg} for the purpose of classifying malware based on their image representations.

\begin{table}[th]
\caption{Relation between File Size and Image Width~\cite{malimg}}
\label{tab:fsiw}
\centering
\begin{tabular}{|l|c|c|c|c|c|c|c|c|}
\hline
\textbf{File Size (in KB)} & <10 & 10 - 30 & 30 - 60 & 60 - 100 & 100 - 200 & 200 - 500 & 500 - 1000 & >1000 \\ \hline
\textbf{Width}             & 32   & 64    & 128   & 256    & 384     & 512     & 768      & 1024  \\ \hline
\end{tabular}
\end{table}

\begin{figure}[ht]\center
\begin{tabular}{@{}c@{\ }c@{\ }c@{\ }c@{\ }c@{}}
\fbox{\includegraphics[width=.16\textwidth]{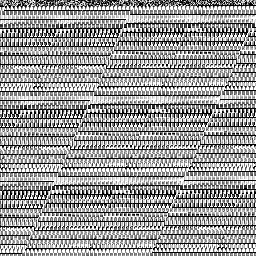}}&
\fbox{\includegraphics[width=.16\textwidth]{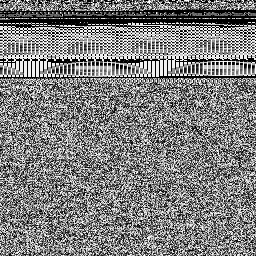}}&
\fbox{\includegraphics[width=.16\textwidth]{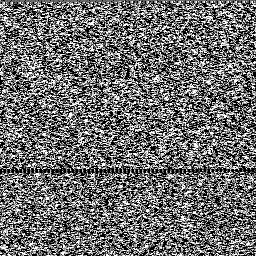}}&
\fbox{\includegraphics[width=.16\textwidth]{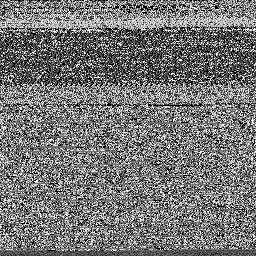}}&
\fbox{\includegraphics[width=.16\textwidth]{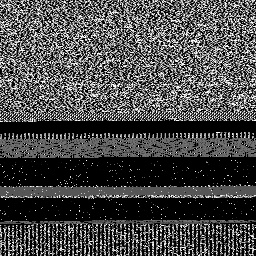}}\\
Lollipop & Kelihos\_ver3  & Tracur & Obfuscator.ACY & Gatak\\
\fbox{\includegraphics[width=.16\textwidth]{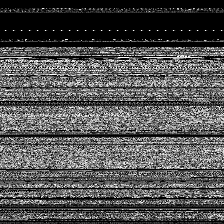}}&
\fbox{\includegraphics[width=.16\textwidth]{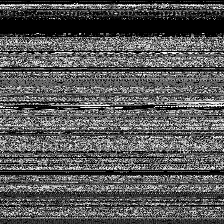}}&
\fbox{\includegraphics[width=.16\textwidth]{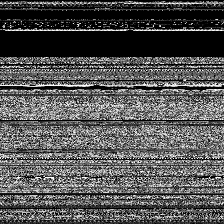}}&
\fbox{\includegraphics[width=.16\textwidth]{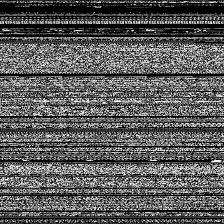}}&
\fbox{\includegraphics[width=.16\textwidth]{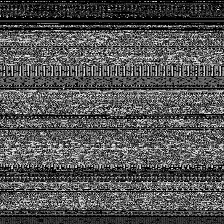}}\\
Adposhel & Amonetize  & Autorun & InstallCore & Benign\\

\end{tabular}
\caption{Examples of constructed images for selected malware families from Microsoft BIG dataset~\cite{ronen2018microsoft} (top part) and Dumpware10~\cite{dumpware10} (bottom part, includes benign sample). }
\label{fig:samples}
\end{figure}

In this work, we apply the byteplot method to transform binary files, both malware and benign, into image files.  
We first convert each byte of the binary file into an unsigned 8-bit integer, ranging from 0 to 255, such that 0 corresponds to black and 255 to white. 
We then reshape this integer array into fixed-width images, whose height depends on the malware or benign file sizes. We pad the images with zeros to make their heights multiples of 32.  
The fixed widths are chosen based on the previous work of Nataraj et al.~\cite{malimg}, as shown in Table~\ref{tab:fsiw}. 
We generate both gray scale and color images from the integer array. 
For gray scale images, each integer value represents the pixel intensity. 
For color images, we use three consecutive integer values to form the RGB components of each pixel. 
We resize all the images to square images, each of dimension $256 \times 256$ for our work. Some sample gray scale images are shown in Fig. \ref{fig:samples}.

\subsection{MALITE: Lightweight Framework for Malware Detection and  Classification} \label{sub-sec:MALITE}

To perform lightweight malware identification and classification, we present MALITE, a framework that consists of two novel methods. 
The first method, MALITE-HRF, extracts patchwise histogram features from malware and benign images and uses a random forest classifier to distinguish among different malware families. 
The second method, MALITE-MN, leverages a lightweight Convolutional Neural Network (CNN) architecture to learn discriminative features from malware images and perform classification. 
In the subsequent sections, we discuss these methods in detail.
\\
\\
\noindent
\textbf{MALITE-HRF:} 
\begin{figure}[ht]
\centering
\fbox{\includegraphics[width=0.7\textwidth, trim={2mm 10mm 0mm 8mm}, clip]{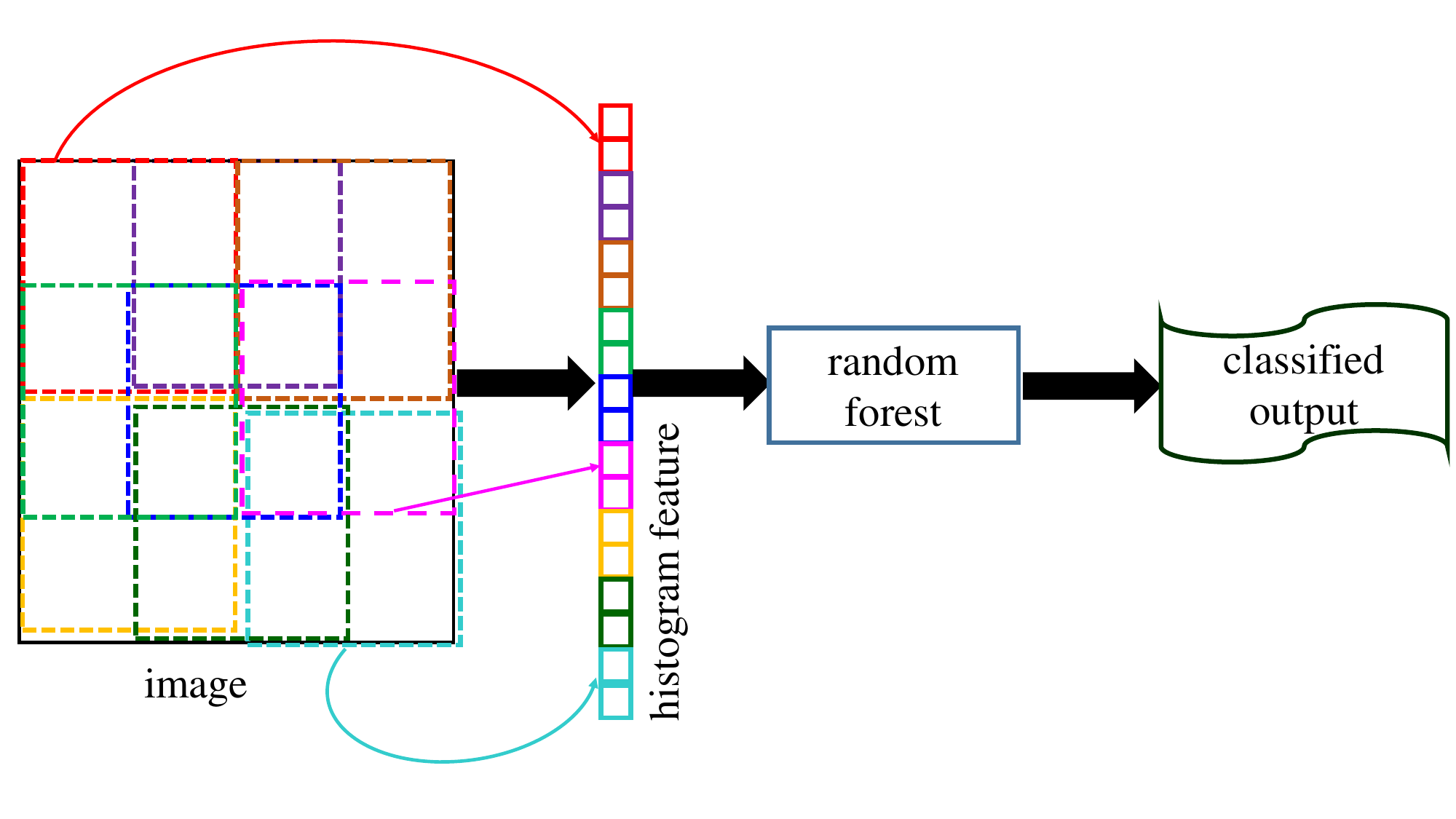}}
\caption{MALITE-HRF: Each colored box denotes a sliding window from which histogram features are extracted that are used for classification using random forest}
\label{fig:malite-hrf}
\end{figure}
A lightweight technique for malware identification and classification, MALITE-HRF is developed by computing a histogram of the intensity values present in a malware or benign image. 
A histogram of the intensity values represents the frequency of a particular intensity value present in an image. 
We utilize this concept to understand the frequency of hex codes present in the original malware and benign binary files. 
Usually, hex codes present in the binary files refer to some assembly language instructions. 
Our hypothesis is that combinations of a few unique sets of such instructions constitute a specific binary file, 
and one binary file differs from another based on the presence of such unique sets of instructions. 

We utilize the concept of the histogram on patches of a binary image to identify the unique sets of instructions. 
The patches in the images are obtained in such a way that the height and the width of each patch are factors of 8. 
We have experimented with different patch sizes, and we empirically found that a patch of dimension $32 \times 256$  for an image of dimension $256 \times 256$ gives the best result when patches are computed with an overlapping window of 50\%. Overall block diagram of MALITE-HRF is shown in Fig.~\ref{fig:malite-hrf}.

\begin{figure}[ht]\center
\begin{tabular}{@{}c@{\ }c@{\ }c@{}}
Ramnit(Sample 1) & Kelihos\_ver3(Sample 1) & Obfuscator.ACY(Sample 1) \\
\fbox{\includegraphics[width=.28\textwidth]{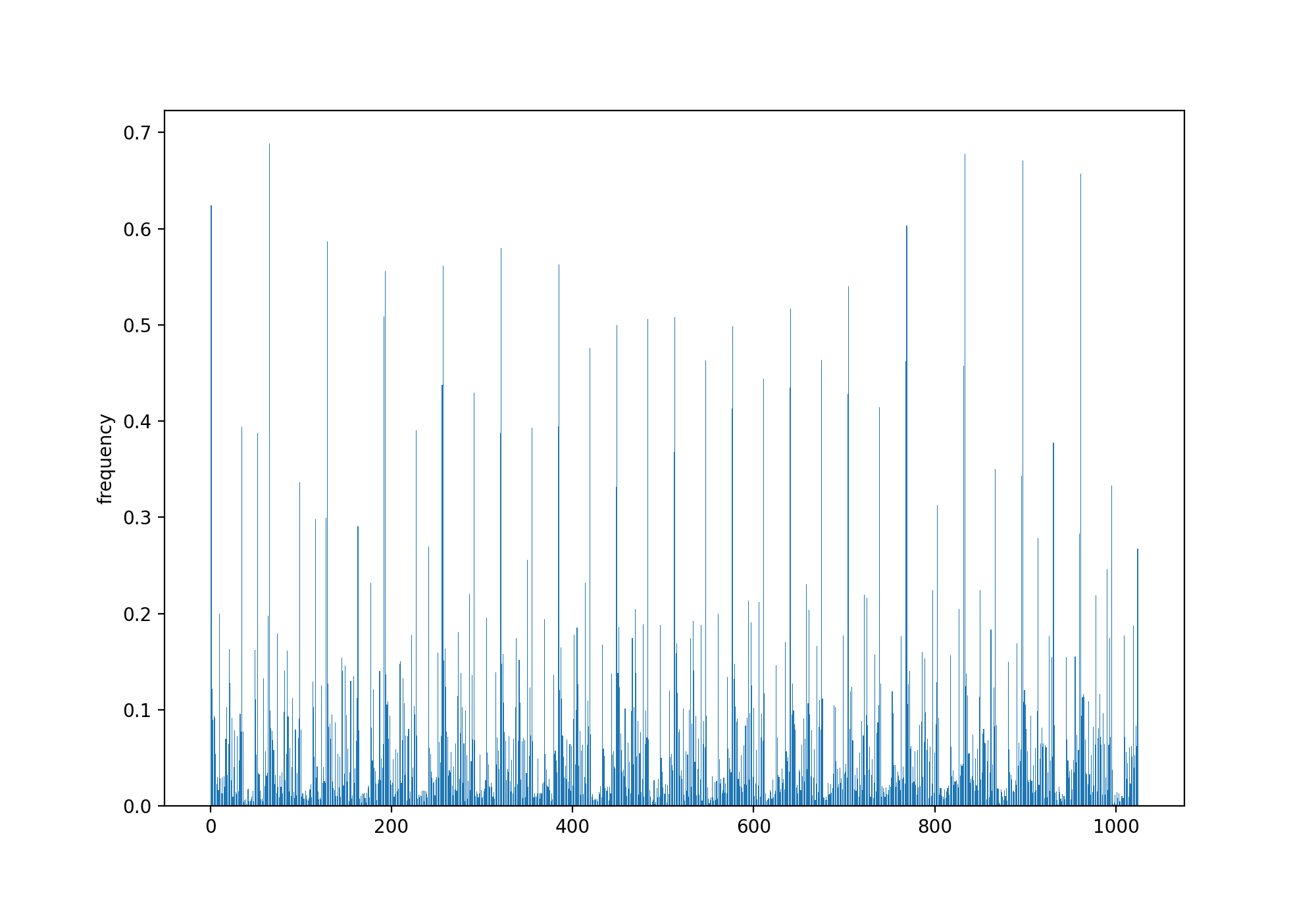}}&
\fbox{\includegraphics[width=.28\textwidth]{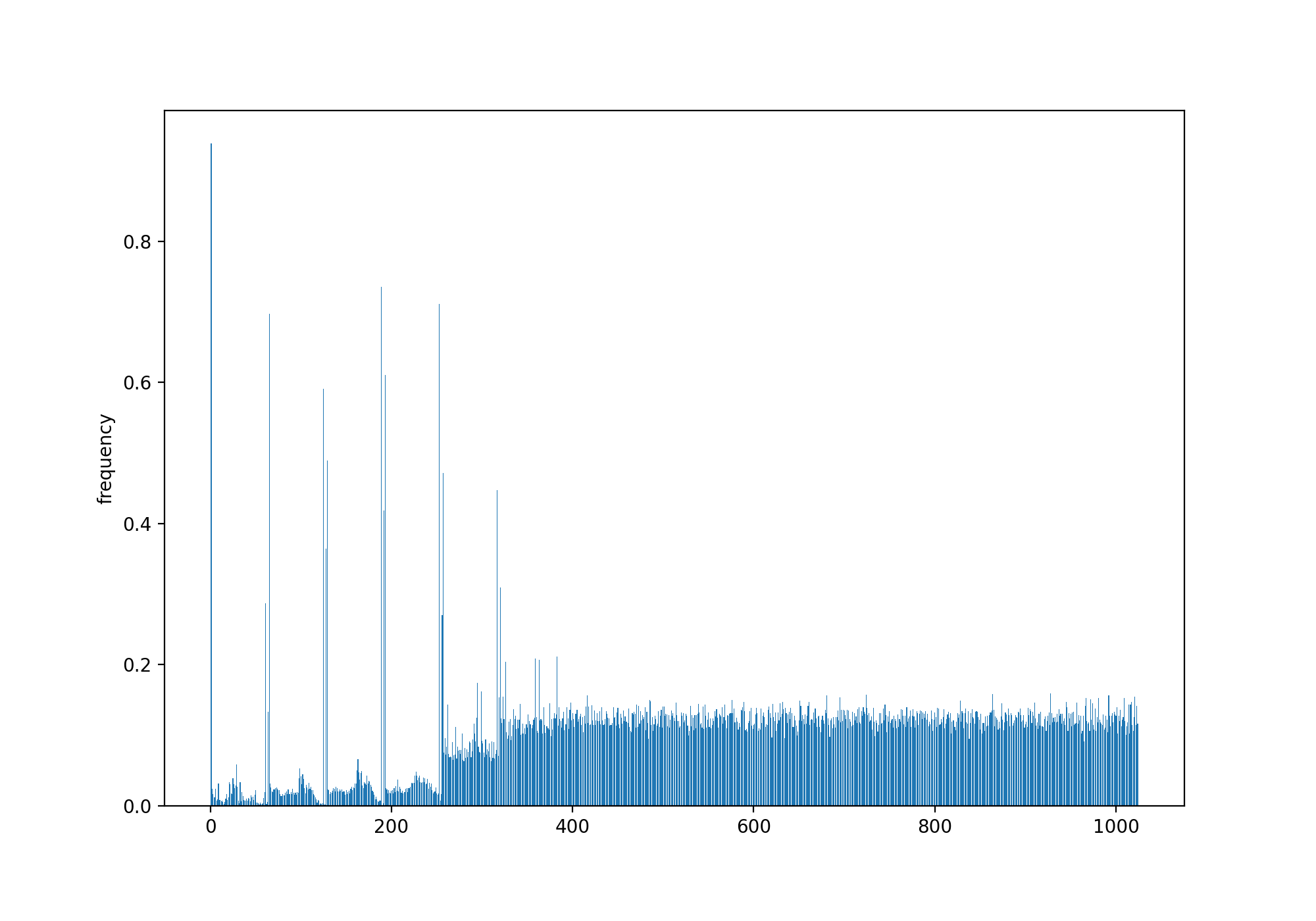}}&
\fbox{\includegraphics[width=.28\textwidth]{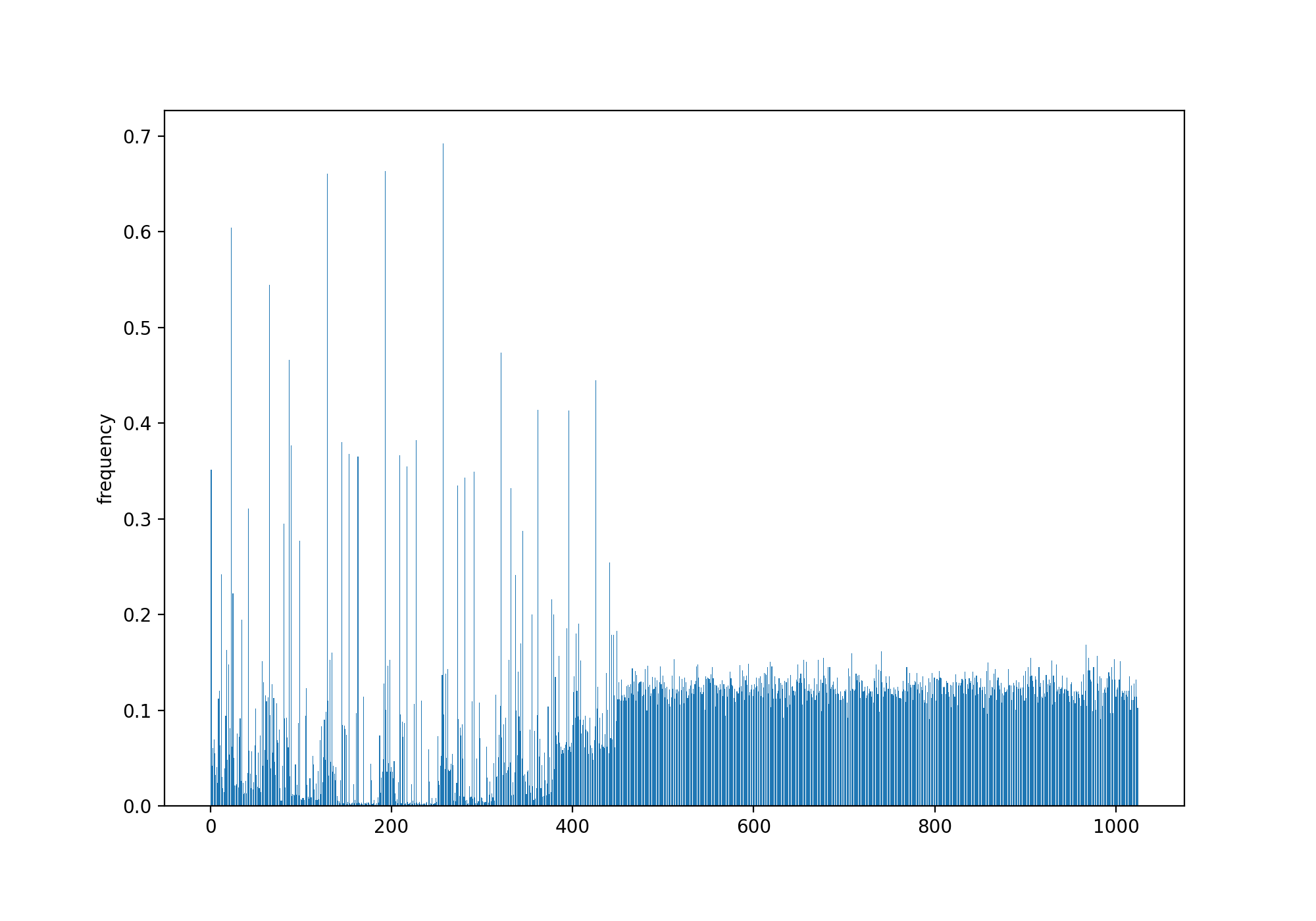}}\\
Ramnit(Sample 2) & Kelihos\_ver3(Sample 2) & Obfuscator.ACY(Sample 2) \\
\fbox{\includegraphics[width=.28\textwidth]{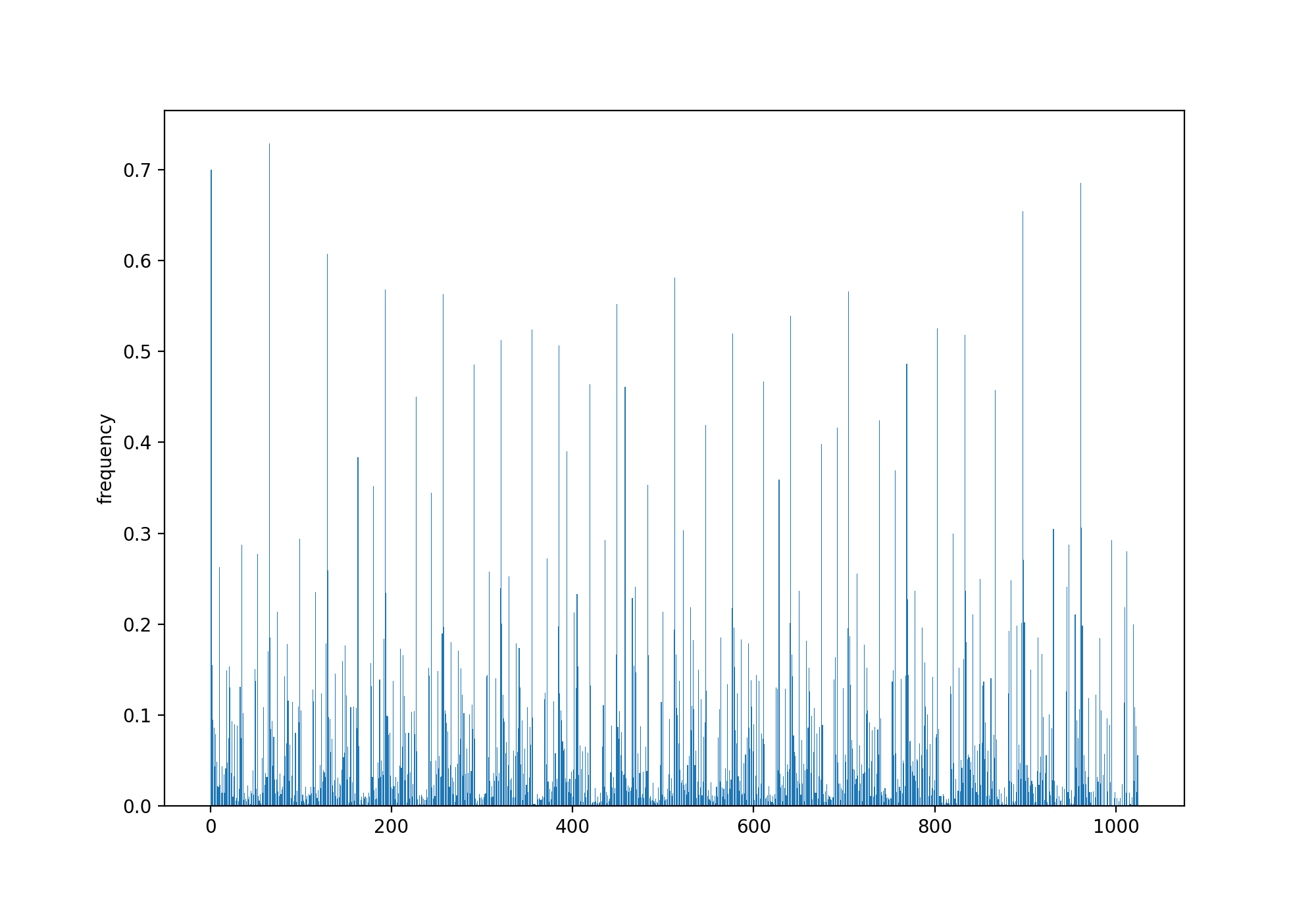}}&
\fbox{\includegraphics[width=.28\textwidth]{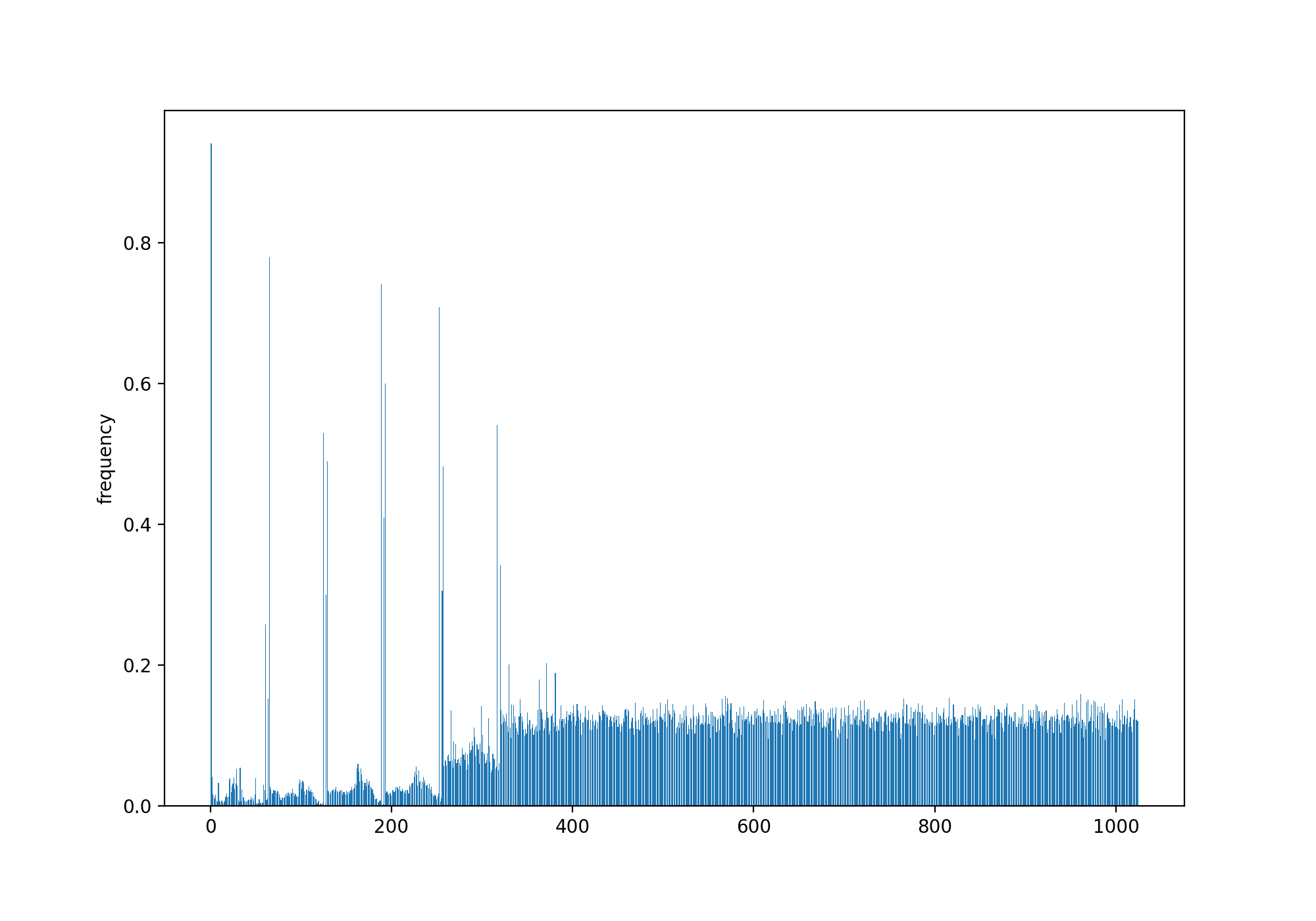}}&
\fbox{\includegraphics[width=.28\textwidth]{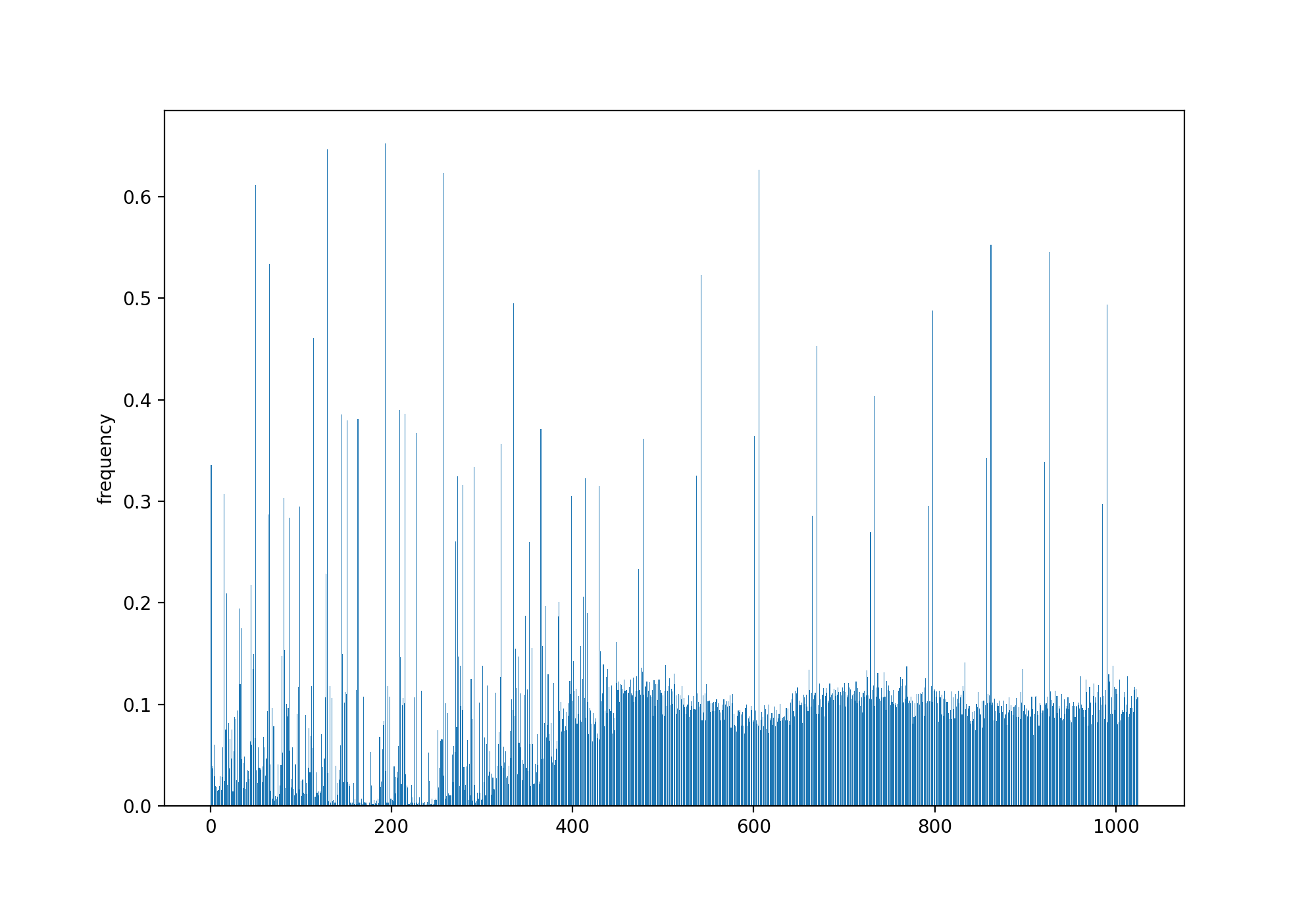}}\\

\end{tabular}
\caption{Typical examples of histogram features computed from Microsoft BIG dataset~\cite{ronen2018microsoft} for 3 randomly selected malware families, namely, Ramnit, Kelihos\_ver3, and Obfuscator.ACY (x-axis: feature dimension, y-axis: frequency). }
\label{fig:hist_examples}
\end{figure}

To reduce the model size and overall computational cost, we further binned the histogram range values into 64 bins empirically. 
Thus, each patch is represented with 64-dimension histogram features, and for an image of dimension $256 \times 256$, we obtain 16 such patches.
Therefore, an image of dimension $256 \times 256$ is represented with a 1024-dimensional feature vector.  
Typical examples of extracted histogram features for a few selected malware families from the Microsoft BIG dataset~\cite{ronen2018microsoft} are shown in Fig.~\ref{fig:hist_examples}. 
From this figure, it is evident that there are indeed some common and distinct patterns present between two binary files from the same and different malware families, respectively. 
We train a random forest classifier with this 1024-dimensional feature vector for the purpose of malware identification and classification process.
\\
\\
\noindent
\textbf{MALITE-MN:} The design choice of MALITE-MN is influenced by the efficient mobile architectures for computer vision applications. MobileNetV2~\cite{mobilenetv2} is one such architecture. 
The key component of MALITE-MN is the residual bottleneck block, which reduces the computational complexity of convolutional layers by a factor of $k^{2}$, where $k$ is the convolution kernel size. 
In our case, we choose $k=3$, resulting in a nine-fold reduction in number of multiplication and addition operations. 
The typical composition and structure of bottleneck layers are shown in Table~\ref{tab:bottleneck}.

\begin{figure}[ht]
\centering
\fbox{\includegraphics[width=0.9\textwidth]{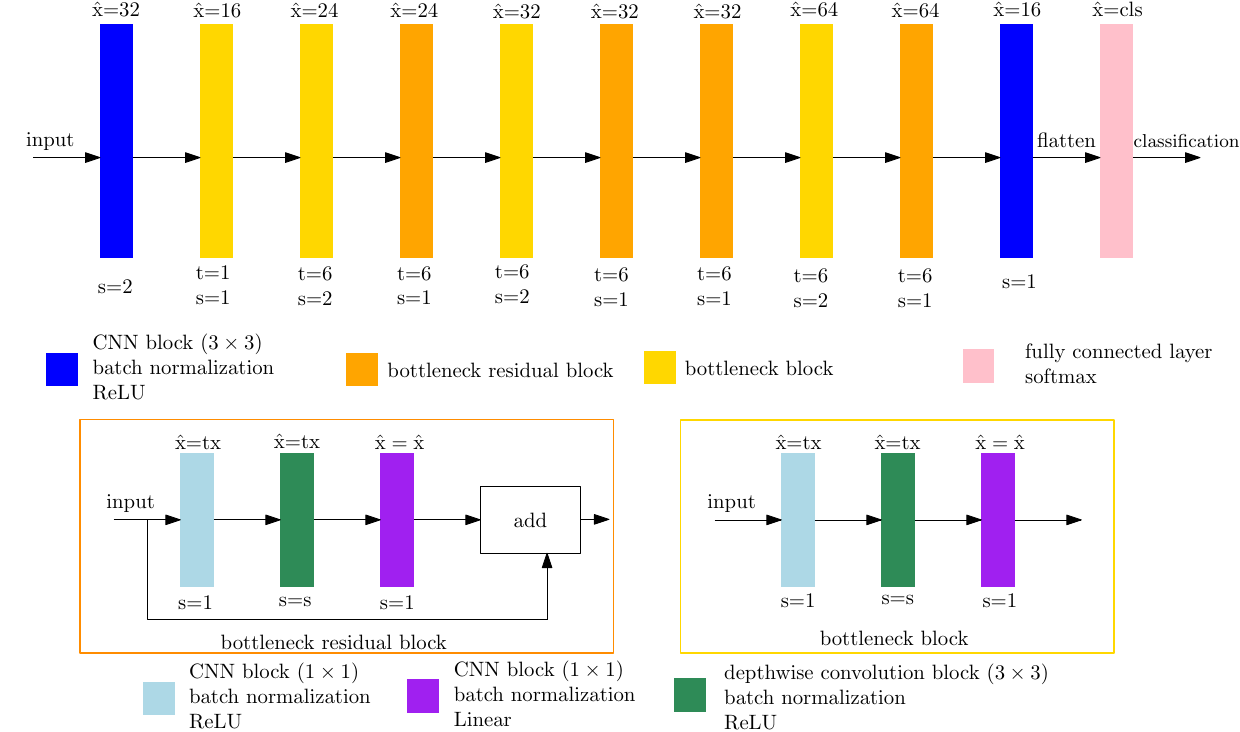}}
\caption{MALITE-MN Architecture. In this figure x, $\hat \textrm{x}$, and s respectively represent input channels, output channels, and stride per layer, whereas, t represents the expansion factor of the bottleneck layer and cls represents the number of target classes for the classification task.}
\label{fig:modelarch}
\end{figure}

We describe the architecture of MALITE-MN as follows.
The image embedding extractor consists of a convolutional layer, followed by eight bottleneck blocks, followed by another convolutional layer.  
We apply batch normalization and activation layers after each layer in the extractor. 
We use a kernel size of $3 \times 3$ for all convolutional and bottleneck layers. 
The expansion factor $t$ of the bottleneck blocks is set to 6, except for the first one, based on the recommendation of Sandler et al.~\cite{mobilenetv2}. 
The classification head consists of a fully connected layer with a softmax activation layer. 
The number of output channels of the fully connected layer matches the number of target classes of the classifier. The overall architecture of MALITE-MN is shown in Fig. \ref{fig:modelarch}.
\begin{table}[h]
\caption{Bottleneck residual block~\cite{mobilenetv2}, where $t$ is the expansion factor, $x$ and $\hat \textrm{$x$}$ are input and output channels of the bottleneck layer, $s$ is the stride size. conv2d ($1 \times 1$), stride=1 refers to a convolution layer with kernel size of ($1 \times 1$), dconv ($3\times 3$), stride=$s$ refers to a depthwise separable convolution with kernel size of ($3 \times 3$) and stride $s$, whereas BN refers to batch normalization layer.}
\label{tab:bottleneck}
\centering

\begin{tabular}{|l|l|l|}
\hline
\multicolumn{1}{|c|}{\textbf{Input Resolution}} & \multicolumn{1}{c|}{\textbf{Layers}}              & \multicolumn{1}{c|}{\textbf{Output Resolution}}     \\ \hline
$h \times w \times x$                  & conv2d ($1\times 1$), stride=1 + BN + ReLU & $h \times w \times tx$                     \\ \hline
$h \times w \times tx$                 & dconv ($3\times 3$), stride=$s$ + BN + ReLU  & $\frac{h}{s} \times \frac{w}{s} \times tx$ \\ \hline
$\frac{h}{s} \times \frac{w}{s} \times tx$ & conv2d ($1\times 1$), stride=1 + BN + Linear & $\frac{h}{s} \times \frac{w}{s} \times \hat x$ \\ \hline
\end{tabular}
\end{table}
\noindent

\section{Results and Discussion}
\label{sec:results}
We present the datasets, experimental setup, and results of our experiments in this section. 
We evaluate MALITE on different publicly available datasets and analyze its performance in detail.  

\subsection{Dataset Description} 
\label{sub-sec:dataset}

In this sub-section, we provide a brief description of the open-source datasets that we have used for our experiments.\\
\textbf{Malimg \cite{malimg}}: This dataset contains 25 malware families and has an overall sample count of 9,458. Some malware families present are Allaple.L, Yuner.A, Lolyda.AA 1, Instantaccess, Fakerean, Adialer.C, Dontovo.A, Skintrim.N. The dataset contains gray scale image representations of the different malware binaries.\\
\textbf{Microsoft BIG \cite{ronen2018microsoft}}: This dataset was published in 2015 with the inception of the Microsoft Malware Classification Challenge. The dataset is almost 0.5 TB in uncompressed form. It contains 9 families of malware such as Gatak, Lollipop, Vundo, Ramnit, Simda, Obfuscator.ACY, Kelihos\_ver1, Kelihos\_ver3 and Tracur.\\
\textbf{Dumpware10 \cite{dumpware10}}: Dumpware10 contains 11 classes out of which one is the benign class and 10 are malware families. It contains a total of 4,294 samples that include 3,433 for training and 861 for validation. The samples are present in the form of RGB images. The malware families included in this dataset are Adposhel,  BrowseFox, Allaple, Dinwod, Amonetize, InstallCore, AutoRun, VBA, MultiPlug and Vilsel.\\
\textbf{MOTIF \cite{motif}}: MOTIF contains 3,095 malicious samples spanning across 454 malware families. To the best of our knowledge, this is the largest open-source malware dataset till date. The dataset was labelled using the publicly available threat reports of several cyber security organizations. However, the dataset is heavily imbalanced due to the presence of a large number of classes. In our experiments, we have considered classes having more than 5 samples. The total number of such classes is 136. \\
\textbf{Drebin \cite{drebin}}: The Drebin dataset is an android malware dataset and was published in 2014. It contains 5,560 malware samples from 179 families of malware. The top 20 classes of this dataset include FakeInstaller, Opfake, Adrd, GingerMaster, Kmin,  Plankton, Geinimi, DroidDream, FakeRun, Gappusin, MobileTx, LinuxLotoor, Iconosys, BaseBridge, DroidKungFu, SMSreg, GoldDream, FakeDoc, SendPay and Imlog and contain a total of 1,048 samples. In our experiments, we have used these top 20 classes.\\
\textbf{CICAndMal2017 \cite{cic_mal_2017}}: This android malware dataset consists of 429 malware samples and 5,065 benign samples. It contains samples from Adware, Ransomware, Scareware and SMS malware spanning 42 malware families some of which are Dowgin, Gooligan, Jisut, Pletor, AVpass, FakeTaoBao, Biige, Mazarbot.

\subsection{Experimental Setup}
\label{sub-sec:ES}


\begin{figure}[t]
\centering
\includegraphics[width=0.75\textwidth]{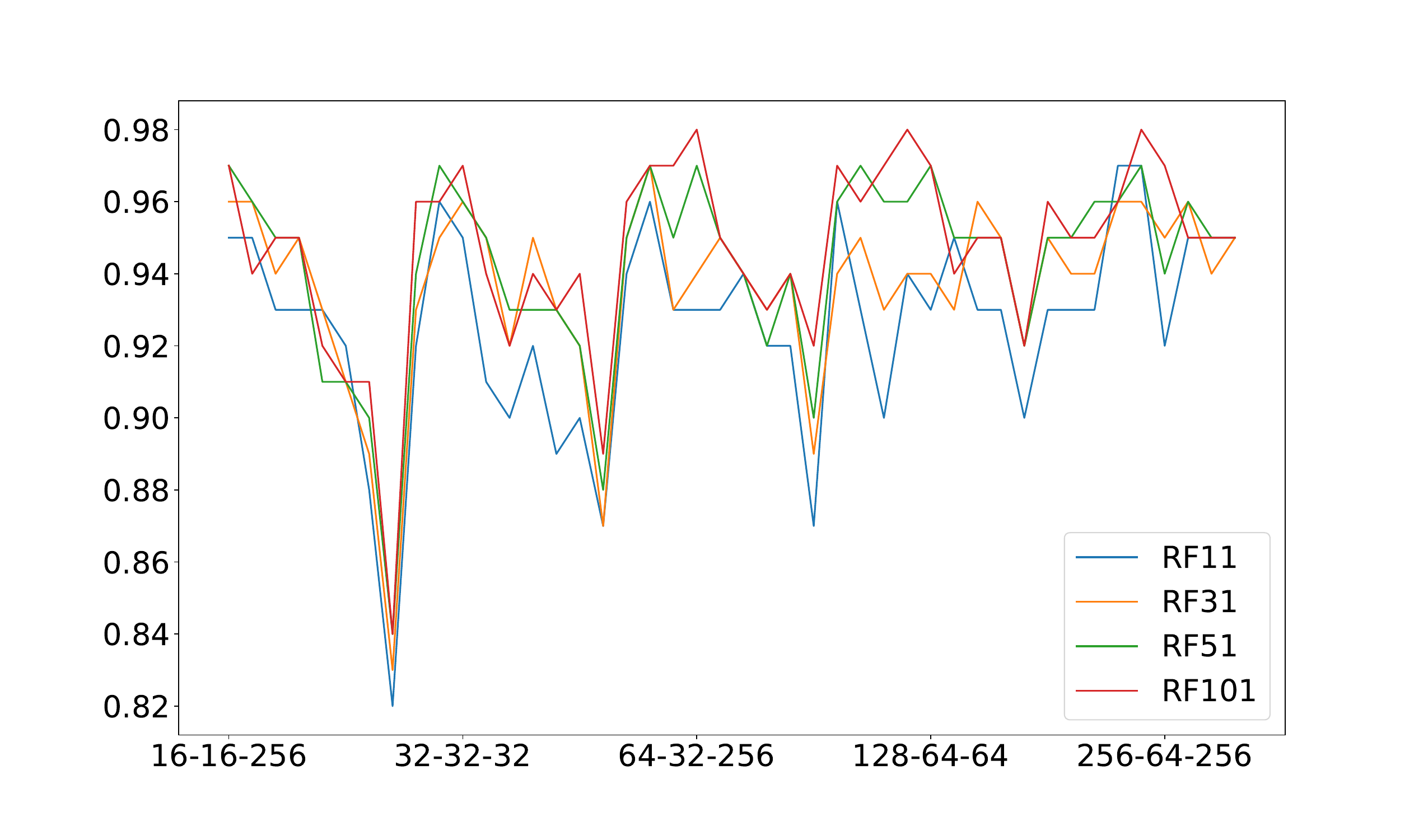}
\caption{Parameter selection for MALITE-HRF using precision as an evaluation metric. Along x-axis, we present various experimental setups using bin size of the histogram, height and width of the patch for feature computation. The x-axis is labeled in the format of `bin-ph-pw', where bin corresponds to bin size and ph and pw represent the height and width of a patch, respectively. The y-axis denotes precision.}
\label{fig:experiments}
\end{figure}

The experiments of MALITE-HRF involved varying several parameters for feature extraction: the bin size of the histogram, the patch height and width, and the number of estimators in the random forest classifier. 
Each experiment is labeled as `bin-ph-pw', where bin, ph, and pw correspond to the parameters bin size, patch height and patch width respectively. 
The patch height and width were constrained by ph $\leq$ pw, and pw was either equal to ph or 256. 
The patch height values were set to 8, 16, 32, 64, 128, and 256. The bin size values ranged from 16 to 256. 
The random forest classifier was tested with four different numbers of estimators: 11, 31, 51, and 101, all with a maximum depth of 15. 
The Microsoft BIG dataset \cite{ronen2018microsoft} was used for the experiments for parameter selection. 
Fig.~\ref{fig:experiments} presents the results. The legend for each line graph is labeled as RFe, where e denotes the number of estimators used for random forest. The figure shows that the optimal parameters are bin = 64, ph = 32, pw = 256, and 51 estimators. 
The performance improved with increasing bin size from 16 to 64, but not further. The number of estimators affected the trade-off between accuracy and model size.


In MALITE-MN model, we varied the expansion factor $t$ of the bottleneck layers from 5 to 10. 
We found that the performance of the model was not sensitive to the choice of $t$ within this range for various vision classification tasks.
Therefore, we fixed $t$ at 6 for the rest of the experiments, following the suggestion of Sandler et al.~\cite{mobilenetv2}. 

To analyze MALITE with respect to various malware classification and identification tasks on the aforementioned datasets, we have used four state-of-the-art approaches for malware analysis. 
Details of the models used in our experiments are as follows: \\
\textbf{3C2D:} The model was proposed by Mohammed et al.~\cite{3c2d} for malware classification. 
This model is a simple CNN based model with three convolutional layers followed by two fully connected layers, and thus the name 3C2D. \\
\textbf{DTMIC:} Kumar et al.~\cite{KUMAR2022103063} proposed a transfer learning based method for malware classification. 
In this model, they had used VGG16~\cite{VGG16} network pretrained on Imagenet dataset~\cite{imagenet_cvpr09}. 
They had frozen the initial encoder part of the network for training. 
During training, only the classification head of the network was trained for malware classification purposes. \\
\textbf{SDN-LSVM:} Wong et al.~\cite{9631209} proposed a transfer learning based method that leveraged two pretrained models, ShuffleNet~\cite{shufflenet} and DenseNet-201~\cite{densenet}, for feature extraction. These models, with 173 and 201 layers respectively, were trained on the ImageNet dataset~\cite{imagenet_cvpr09}. 
They concatenated the feature vectors obtained from the global average pooling layer of each model and fed them to a linear SVM classifier with a one vs. one scheme. In the rest of the paper, we refer to this approach as SDN-LSVM.\\
\textbf{MalConv2:} The memory-improved version of the MAlConv model was proposed by Raff et al.~\cite{malconv} for the classification of malware by consuming the whole malware binary file as sequential data.  
MalConv2 consists of two 1D convolutional layers. A 1D convolutional layer uses a filter and a stride, each of size 512, 128 channels and 8 embedding dimensions. \\
We trained all the models using the Adam optimizer and the Categorical Cross-Entropy Loss. 
We applied cosine annealing with a quarter period to the learning rate, which decayed from $1e^{-4}$ to $5e^{-5}$ for 1000 epochs, with a warmup phase of $5000$ steps. 
The training was performed on an Nvidia A100 GPU.

\subsection{Results and Discussion}
\label{sub-sec:rd}
Table~\ref{tab:complexity} compares the proposed models MALITE-HRF and MALITE-MN with four existing state-of-the-art approaches for malware detection and categorization based on the number of parameters, the number of multiplication-addition (Mult-Add) operations, and the size of the models. 
The number of parameters reflects the complexity and the capacity of the models. The number of Mult-Add operations is directly proportional to the model computational complexity which in turn relates to the inference time and the battery power consumption of the device on which the model is deployed.
The size of a model represents its memory requirement and storage overhead.
In our work, we compute the number of multiplication and addition operations for convolutional layers and bottleneck layers by using the methods described in~\cite{mobilenetv2}. For histogram feature extraction, the number of operations is equivalent to ($\text{n} \times \text{ph} \times \text{pw}$), where n is the number of patches, ph and pw are the height and width of each patch, respectively. For random forest, the number of computations is atmost ($\text{e}\times\text{ht}$), where e is the number of estimators present in the random forest classifier and ht is the height of each such estimator tree. 

In Table~\ref{tab:complexity}, all the values are computed considering input binary images of dimension 256 $\times$ 256, and the number of output classes to be 10. In case of Malconv2, the input image of dimension 256 $\times$ 256 is considered as a single dimensional array.
\begin{table}[tb]
\caption{\# Parameters, \# Multiplication-Addition Operations, and Sizes of Models.}
\begin{center}
\footnotesize
\setlength{\tabcolsep}{3pt}
\begin{tabular}{@{}lS[table-format=2.2]S[table-format=5.2]S[table-format=3.2]@{}} 
\hline
\multirow{2}{*}{\textbf{Model}} &   {\textbf{\# Parameters}} & {\textbf{\# Mult-Adds Ops.}} & {\textbf{Size}}  \\ 
    &       {\textbf{(in millions)}} & {\textbf{(in millions)}} & {\textbf{(in MB)}}  \\ 
\hline    
         \rowcolor[HTML]{E6E9ED}
         MALITE-HRF (proposed) & 0.01 & 0.13 & 0.03 \\ 
         \rowcolor[HTML]{E6E9ED}
         MALITE-MN (proposed) & 0.18 & 303.54 & 0.81 \\ 
         3C2D~\cite{3c2d} & 67.61 & 727.85 & 276.46 \\ 
         DTMIC~\cite{KUMAR2022103063} & 17.92 & 15353.06 & 71.74 \\ 
         SDN-LSVM~\cite{9631209} & 23.27 & 18724.06 & 82.96 \\ 
         MalConv2~\cite{malconv} & 1.07 & 68719.51 & 4.30 \\ \hline
    \end{tabular}
\end{center}
\label{tab:complexity}
\end{table}

From the table, it can be seen that the proposed methods use a significantly lower number of parameters, Mult-Add operations, and have a significantly lesser size than the existing approaches. This implies that our techniques are more efficient and lightweight for malware detection. 
Specifically, MALITE-HRF employs only 0.01 million parameters, 0.13 million Mult-Add operations, and has a size of 0.03 MB, which are orders of magnitude lower than the other models, thus making MALITE-HRF ultra lighweight. 
MALITE-MN uses 0.18 million parameters, 303.54 million Mult-Add operations, and has 0.81 MB size, which are also several times lower than 3C2D, DTMIC, SDN-LSVM and MalConv2. 

The existing methods, on the other hand, have a much higher number of parameter count, Mult-Add operations, and size than the proposed models, which indicate that they are more complex and resource intensive for malware detection. 
Among them, 3C2D has the highest number of parameters (67.61 million) and the largest size (276.46 MB), which are mainly due to the use of high-dimensional fully connected layers. 
DTMIC and SDN-LSVM have the higher number of Mult-Add operations (15353.06 million and 18724.06 million, respectively), which can be attributed to their use of deep pretrained CNN networks for feature computation. 
MalConv2 has a moderate number of parameters (1.07 million) and size (4.30 MB), but a very high number of Mult-Add operations (68719.51 million) because of the use of a long 1D convolutional neural network that requires a large number of sliding windows and feature maps. 
Our proposed approaches demonstrate a clear advantage over existing models in terms of efficiency and lightweightness.
It is to be noted here that MALITE is 6,761 (HRF) to 375 (MN) times and 5,598 (HRF) to 2 (MN) times smaller with respect to parameter count and number of Mult-Add operations, respectively than the largest model 3C2D. 
Moreover, MALITE is also smaller than the smallest model MalConv2 with respect to parameter count and Mult-Add operations by 107 (HRF) to 6 (MN) times and 5,28,611 (HRF) to 226 (MN) times, respectively. 

\begin{figure}[h!]\center
\begin{tabular}{@{}c@{\ }c@{}}
\hline
\multicolumn{2}{c}{Malimg} \\ \hline
\includegraphics[width=.45\textwidth]{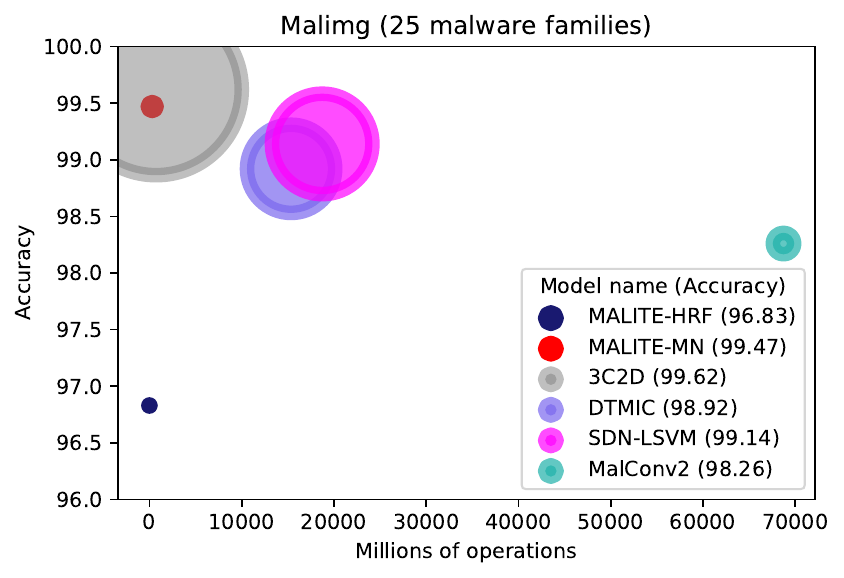}&
\includegraphics[width=.44\textwidth]{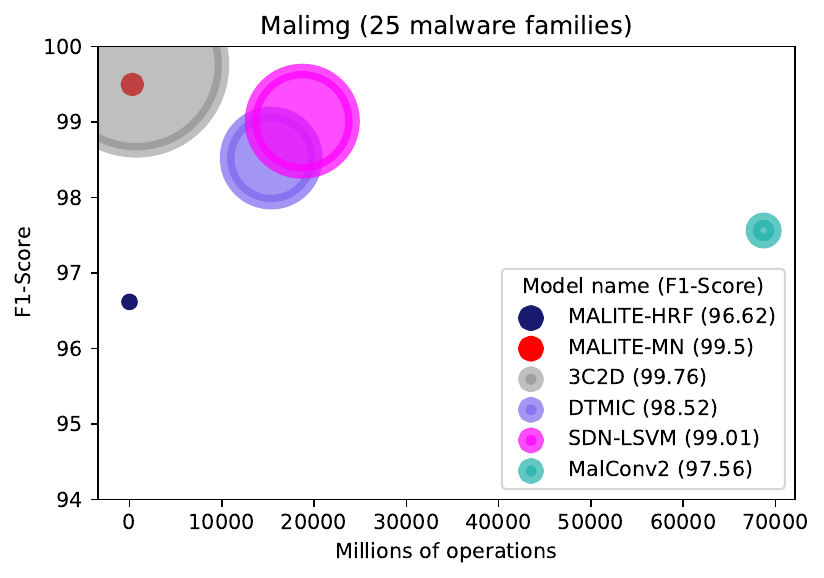}\\
\hline
\multicolumn{2}{c}{Microsoft BIG} \\ \hline
\includegraphics[width=.45\textwidth]{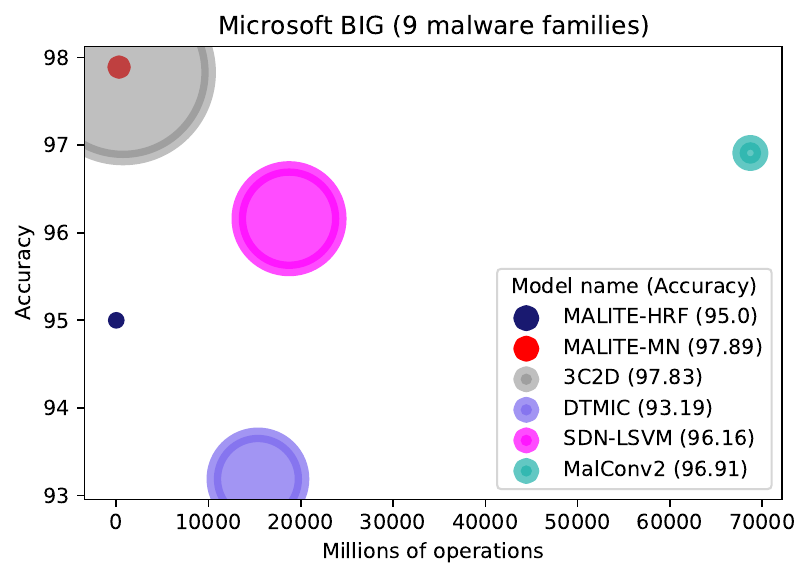}&
\includegraphics[width=.45\textwidth]{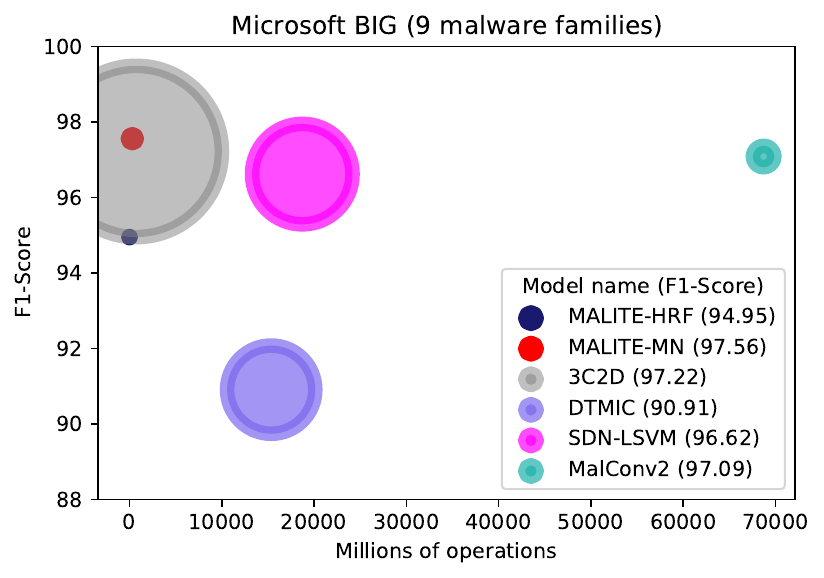}\\
\hline
\multicolumn{2}{c}{Dumpware10} \\ \hline
\includegraphics[width=.45\textwidth]{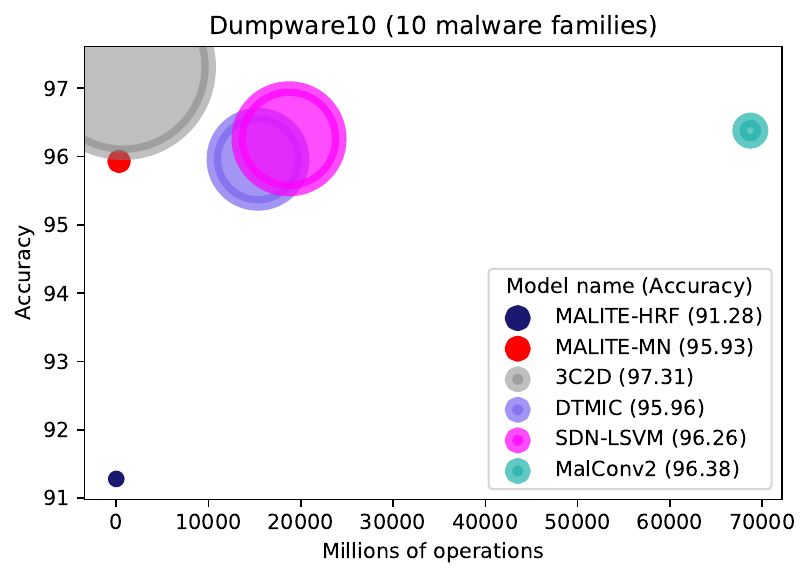}&
\includegraphics[width=.45\textwidth]{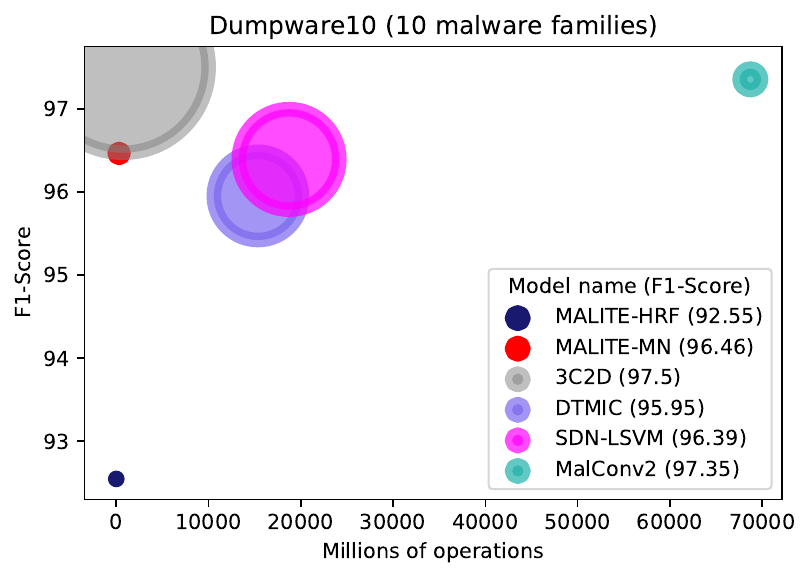}\\
\hline
\multicolumn{2}{c}{MOTIF} \\ \hline
\includegraphics[width=.45\textwidth]{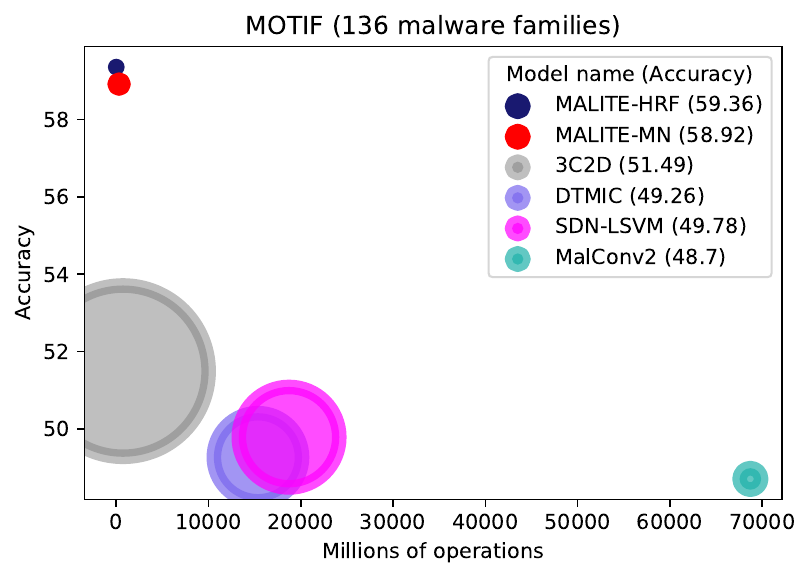}&
\includegraphics[width=.45\textwidth]{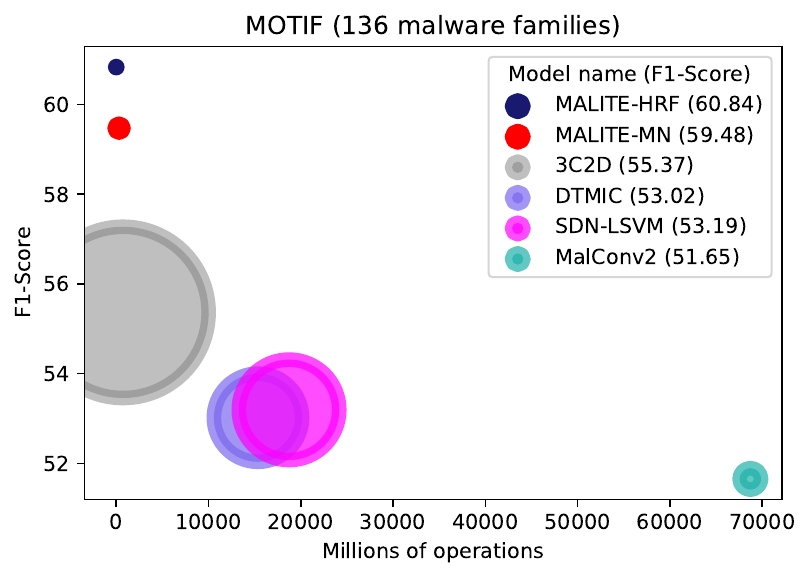}\\
\end{tabular}
\caption{Malware family classification comparison of MALITE with 3C2D, DTMIC, SDN-LSVM and MalConv2 for 4 malware datasets, Malimg, Microsoft BIG, Dumpware10 and MOTIF in terms of Accuracy and F1-Score. In each graph, a model is represented by a colored bubble and the bubble size represents the corresponding model size with respect to the size of the smallest model, MALITE-HRF. x-axis: No. of Mult-Add operations in millions, y-axis: Accuracy or F1-Score.}
\label{fig:malwareclf}
\end{figure}

To evaluate the proposed methods in terms of their effectiveness for malware classification tasks, we compare our models with four state-of-the-art techniques on six publicly available datasets. 
The metrics used in our comparison are accuracy which measures the overall correctness and F1-Score which is the harmonic mean of precision and recall. Precision and recall respectively measure the fraction of true positives among predicted positives, and the fraction of true positives among actual positives. The higher the values of these metrics, the better the performance of the model.

\begin{figure}[t]\centering
\begin{tabular}{@{}c@{\ }c@{}}
\hline
\multicolumn{2}{c}{Android Malware Classification} \\ \hline
\includegraphics[width=.42\textwidth]{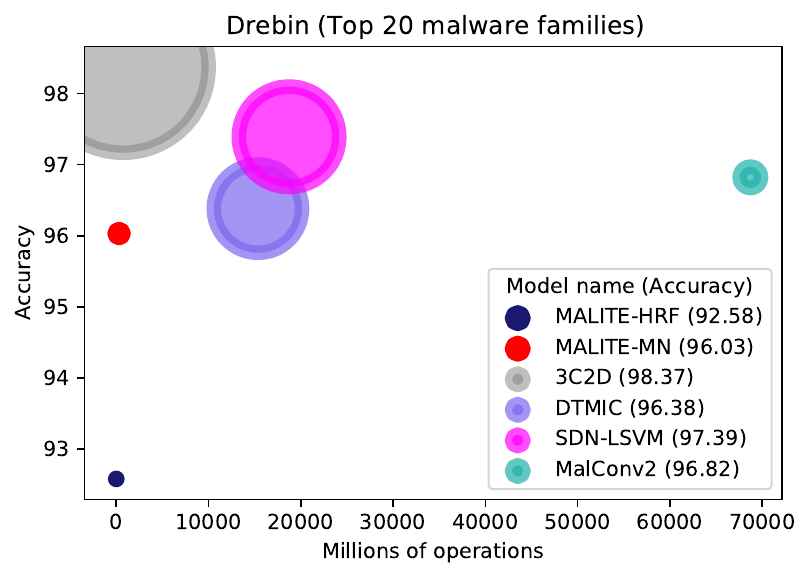}&
\includegraphics[width=.42\textwidth]{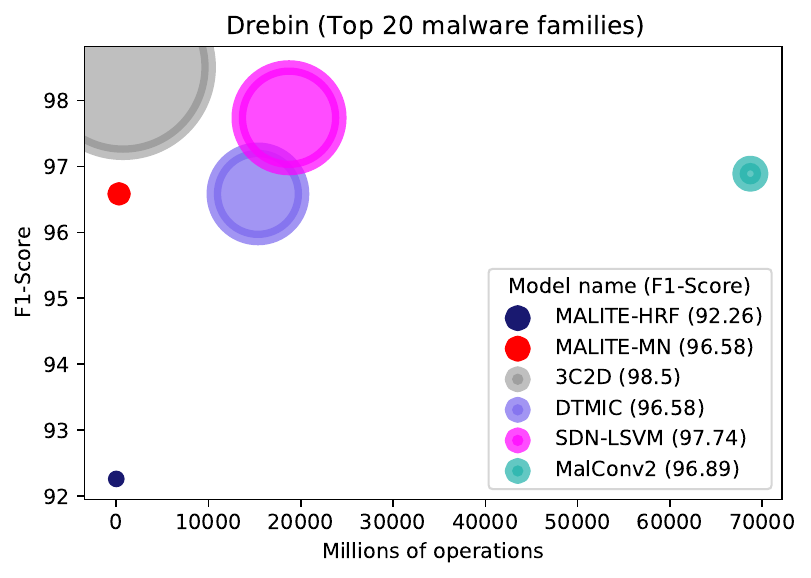}\\
\includegraphics[width=.42\textwidth]{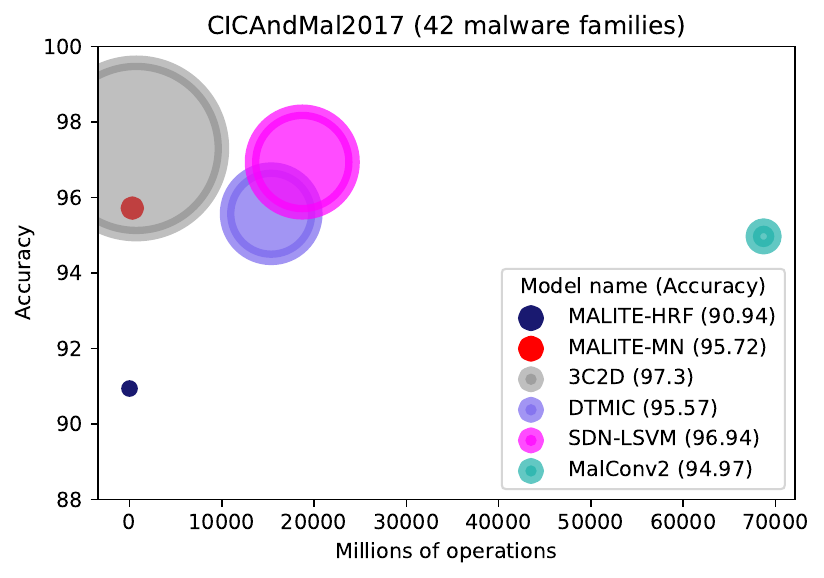}&
\includegraphics[width=.42\textwidth]{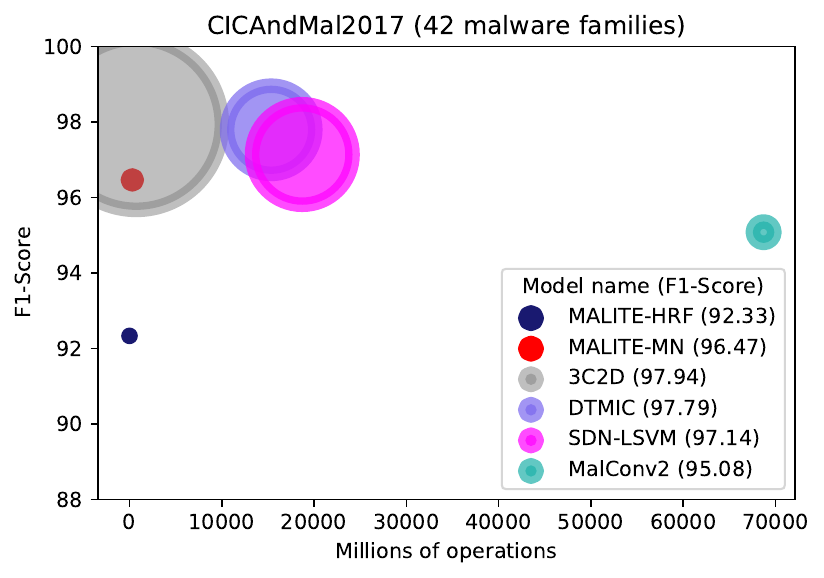}\\
\hline
\multicolumn{2}{c}{Benign vs. Malware Classification} \\ \hline
\includegraphics[width=.42\textwidth]{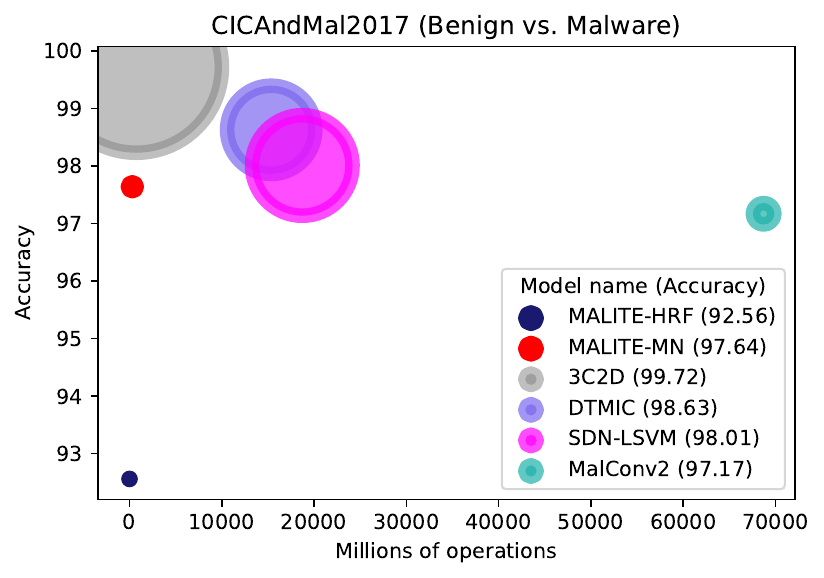}&
\includegraphics[width=.42\textwidth]{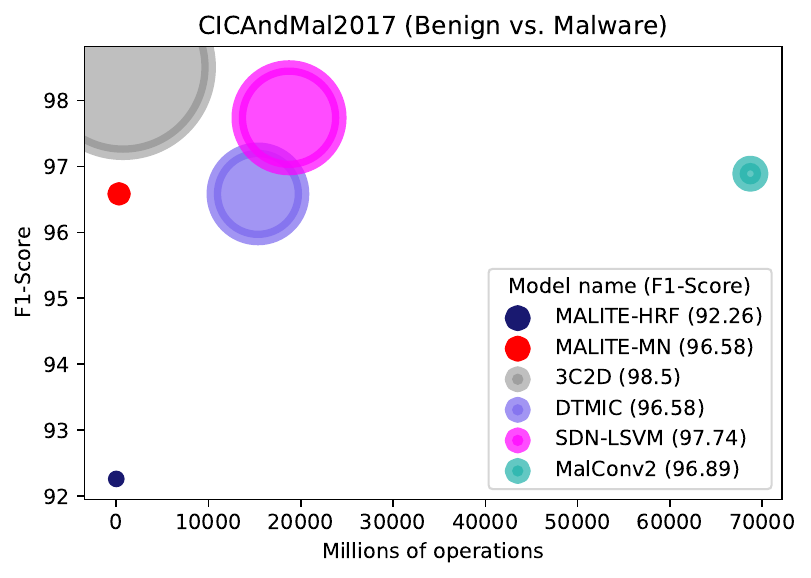}\\
\includegraphics[width=.42\textwidth]{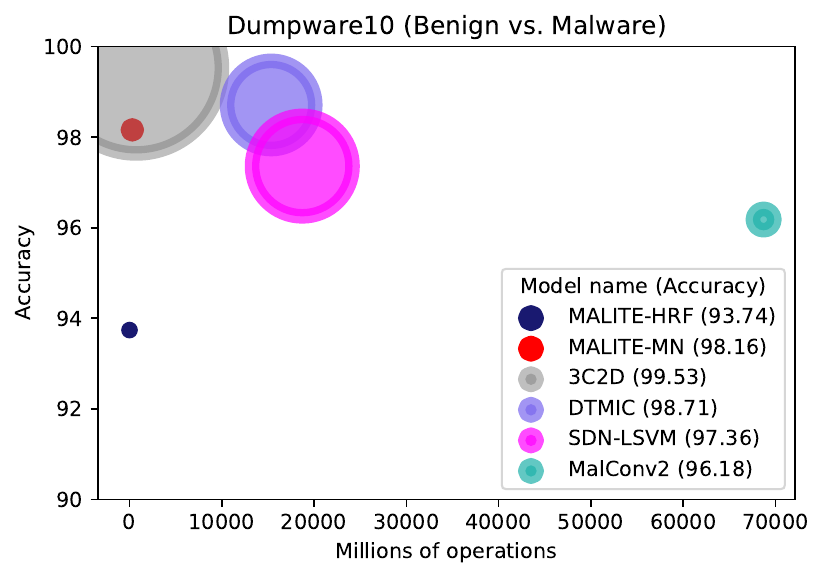}&
\includegraphics[width=.42\textwidth]{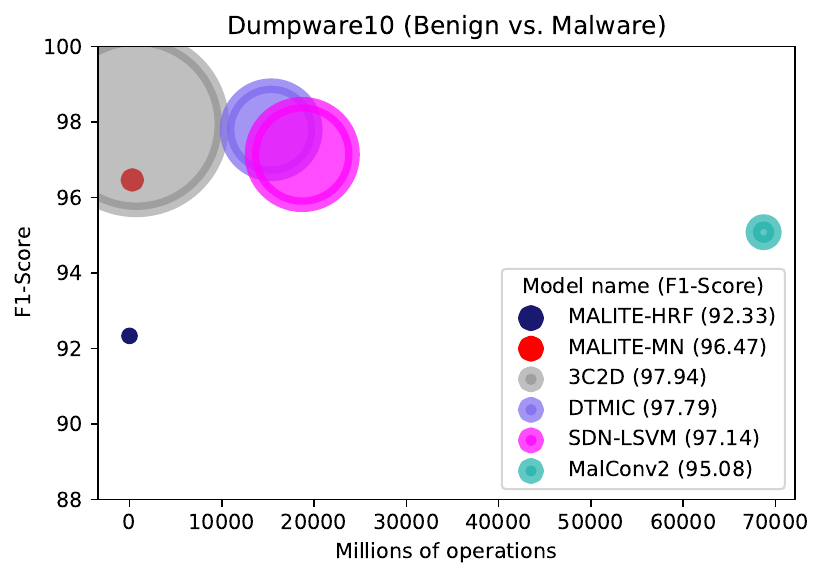}\\

\end{tabular}
\caption{Performance comparison of MALITE with 3C2D, DTMIC, SDN-LSVM and MalConv2. Top figure shows malware family classification results for 2 android malware datasets, Drebin Top20 and CICAndMal2017 and bottom figure shows benign vs. malware classification results for CICAndMal2017 and Dumpware10 in terms of Accuracy and F1-Score. In each graph, a colored bubble represents a model and the bubble size represents the corresponding model size with respect to MALITE-HRF. 
}
\label{fig:malwareclfandroid}
\end{figure}

We present in Fig.~\ref{fig:malwareclf} the results of malware family classification on four datasets namely, Malimg, Microsoft BIG, Dumpware10, and MOTIF using six different models, including our models, MALITE-HRF and MALITE-MN. 
From Fig.~\ref{fig:malwareclf}, it can be observed that MALITE-HRF and MALITE-MN achieve comparable or superior performance than the other four techniques on most of the datasets. MALITE-MN outperforms all the other models on Microsoft BIG, achieving the highest accuracy and F1-Score of 97.89\% and 97.56\% respectively. MALITE-MN gives an accuracy of 99.47\% and an F1-Score of 99.50\% on Malimg which is approximately within 0.25\% of the best performing model 3C2D. MALITE-HRF also performs well on these two datasets, with accuracy and F1-Score values almost equal to or above 95\%. 

On Dumpware10 dataset, the proposed models also show good performance, with accuracy value above 91\% and F1-Score above 92\% for MALITE-HRF and accuracy and F1-Score values of $\approx$96\% for MALITE-MN. However, on this dataset, the 3C2D model achieves the best performance, with accuracy and F1-Score of 97.31\%, and 97.50\%, respectively. This suggests that 3C2D can better handle the diversity and complexity of Dumpware10, which contains malware samples from 10 different sources. However, the performance of MALITE-MN is approximately within 1.4\% of 3C2D.
On the MOTIF dataset, our proposed models outperform all other models by considerable margins of 10\% and 5\% in terms of accuracy and F1-Score, respectively. 
The MOTIF dataset poses significant challenges for malware classification using machine learning and especially, Deep Neural Network (DNN) methods.
First, the dataset is highly imbalanced and it consists of 136 malware families with more than 5 samples. 
Second, the majority of the malware classes have very few data points, which limits the ability of DNN methods to learn meaningful features and generalize well. Here, MALITE-HRF gives the best performance. 

We also compare MALITE on two android malware datasets, Drebin, and CICAndMal2017. 
The comparison of the two proposed methods and the four state-of-the-art approaches is shown in Fig.~\ref{fig:malwareclfandroid}. 
\clearpage 
It can be observed that on both the datasets, 3C2D outperforms all other models with respect to accuracy and F1-Score with values greater than 97\% and $\approx $98\%, respectively. 
However, on these datasets MALITE-MN obtains an accuracy and F1-Score of $\approx$96\% or higher. 
Moreover, we observe that accuracy of MALITE-HRF is approximately equal to or more than 90\% and F1-Score of $\approx$92\%. Thus, at least one of our models give a performance of within 2\% margin of the best model, 3C2D. Overall, the results indicate that the proposed models can effectively capture the structural and semantic features of the malware binaries and distinguish among different malware families.

We further analyze MALITE with respect to benign vs. malware classification on CICAndMal2017 and Dumpware10 since only these two datasets contain benign samples. Performance of the methods is shown in Fig.~\ref{fig:malwareclfandroid} (bottom part). 
It can be observed in this figure that our proposed methods MALITE-HRF and MALITE-MN achieve comparable results on these two datasets for malware vs. benign classification, outperforming some of the baseline models and being close to the best-performing ones. 

Our experimental study demonstrates that our proposed models are capable of performing malware classification and identification with quite a high accuracy, despite being several orders of magnitude smaller than state-of-the-art methods in terms of memory and computational cost. Out of the eight sets of experiments, in terms of F1-Score, MALITE gives best performance in two cases and achieves F1-Score values within less than 0.5\%, 1.5\% and 2\% of the best performing model in one, three and two cases respectively. 
Therefore, we recommend MALITE-MN for resource-constrained devices, while MALITE-HRF is recommended for devices having extreme memory and computation constraints.

\vspace{-0.1in}
\section{Conclusion} \label{sec:conclusion}
\vspace{-0.1in}
In this paper, we have proposed MALITE, a lightweight framework for malware identification and classification. We have designed two variants of MALITE, MALITE-HRF that combines lightweight methods like patch based histogram computation and random forest classifier, and MALITE-MN which is a light weight neural network based classifier using computationally inexpensive bottleneck layers. Experimental results on six open-source datasets demonstrate the effectiveness of the proposed techniques inspite of being several times lighter in terms of computational overhead and memory consumption when compared to state-of-the-art malware analysis approaches, thus making MALITE suitable for constrained computing systems. In future, we intend to design lightweight non-image based malware detection and classification strategies. Moreover, we would also like to design lightweight techniques that are resilient to malware evolution and obfuscation and are capable of detecting zero-day attacks. 

\vspace{-0.1in}
\section*{Acknowledgments}
The authors gratefully acknowledge the computing time provided on the high performance computing facility, Sharanga, at the Birla Institute of Technology and Science - Pilani, Hyderabad Campus.

\FloatBarrier
\bibliographystyle{splncs04}
\bibliography{bibtex/bib/references}

\begin{thebibliography}{10}
\providecommand{\url}[1]{\texttt{#1}}
\providecommand{\urlprefix}{URL }
\providecommand{\doi}[1]{https://doi.org/#1}

\bibitem{drebin}
Arp, D., Spreitzenbarth, M., Hubner, M., Gascon, H., Rieck, K.: Drebin:
  Effective and explainable detection of android malware in your pocket. In:
  NDSS. The Internet Society (2014)

\bibitem{9455368}
Aslan, O., Yilmaz, A.A.: A new malware classification framework based on deep
  learning algorithm. IEEE Access  \textbf{9},  87936--87951 (2021)

\bibitem{Bae:Summary}
Bae, S.I., Lee, G.B., Im, E.G.: Ransomware detection using machine learning
  algorithms. Concurrency and Computation: Practice and Experience
  \textbf{32}(18),  e5422 (2020)

\bibitem{Baptista:2019}
Baptista, I., Shiaeles, S., Kolokotronis, N.: A novel malware detection system
  based on machine learning and binary visualization. In: IEEE International
  Conference on Communications Workshops. pp.~1--6 (2019)

\bibitem{dumpware10}
Bozkir, A.S., Tahillioglu, E., Aydos, M., Kara, I.: Catch them alive: A malware
  detection approach through memory forensics, manifold learning and computer
  vision. Computers \& Security  \textbf{103},  102166 (2021)

\bibitem{brengel2021yarix}
Brengel, M., Rossow, C.: Yarix: Scalable yara-based malware intelligence. In:
  USENIX Security Symposium. pp. 3541--3558 (2021)

\bibitem{CHAGANTI2022103306}
Chaganti, R., Ravi, V., Pham, T.D.: Image-based malware representation approach
  with efficientnet convolutional neural networks for effective malware
  classification. Journal of Information Security and Applications
  \textbf{69},  103306 (2022)

\bibitem{Sonicwall}
Conner, B.: 2022 sonicwall cyber threat report (2022)

\bibitem{Conti20081}
Conti, G., Dean, E., Sinda, M., Sangster, B.: Visual reverse engineering of
  binary and data files. In: International Workshop on Visualization for
  Computer Security. p. 1 – 17 (2008)

\bibitem{10.1007/978-3-030-87839-9_4}
Daoudi, N., Samhi, J., Kabore, A.K., Allix, K., Bissyand{\'e}, T.F., Klein, J.:
  Dexray: A simple, yet effective deep learning approach to android malware
  detection based on image representation of bytecode. In: Wang, G., Ciptadi,
  A., Ahmadzadeh, A. (eds.) Deployable Machine Learning for Security Defense.
  pp. 81--106 (2021)

\bibitem{8626828}
Darus, F.M., Salleh, N.A.A., Mohd~Ariffin, A.F.: Android malware detection
  using machine learning on image patterns. In: 2018 Cyber Resilience
  Conference. pp.~1--2 (2018)

\bibitem{imagenet_cvpr09}
Deng, J., Dong, W., Socher, R., Li, L.J., Li, K., Fei-Fei, L.: {ImageNet: A
  Large-Scale Hierarchical Image Database}. In: CVPR09 (2009)

\bibitem{BYTEcodeimage}
Ding, Y., Zhang, X., Hu, J., Xu, W.: Android malware detection method based on
  bytecode image. Journal of Ambient Intelligence and Humanized Computing
  (2020)

\bibitem{DANGELO202026}
D’Angelo, G., Ficco, M., Palmieri, F.: Malware detection in mobile
  environments based on autoencoders and api-images. Journal of Parallel and
  Distributed Computing  \textbf{137},  26--33 (2020)

\bibitem{ELAYAN2021847}
Elayan, O.N., Mustafa, A.M.: Android malware detection using deep learning.
  Procedia Computer Science  \textbf{184},  847--852 (2021)

\bibitem{8955840}
Fang, Y., Gao, Y., Jing, F., Zhang, L.: Android malware familial classification
  based on dex file section features. IEEE Access  \textbf{8},  10614--10627
  (2020)

\bibitem{9136685}
Feng, J., Shen, L., Chen, Z., Wang, Y., Li, H.: A two-layer deep learning
  method for android malware detection using network traffic. IEEE Access
  \textbf{8},  125786--125796 (2020)

\bibitem{9204665}
Feng, R., Chen, S., Xie, X., Meng, G., Lin, S.W., Liu, Y.: A
  performance-sensitive malware detection system using deep learning on mobile
  devices. IEEE Trans. on Information Forensics and Security  \textbf{16},
  1563--1578 (2021)

\bibitem{GAO2021102264}
Gao, H., Cheng, S., Zhang, W.: Gdroid: Android malware detection and
  classification with graph convolutional network. Computers \& Security
  \textbf{106},  102264 (2021)

\bibitem{10.1145/3162625}
Garcia, J., Hammad, M., Malek, S.: Lightweight, obfuscation-resilient detection
  and family identification of android malware. ACM Trans. Softw. Eng.
  Methodol.  \textbf{26}(3) (2018)

\bibitem{9185490}
Go, J.H., Jan, T., Mohanty, M., Patel, O.P., Puthal, D., Prasad, M.:
  Visualization approach for malware classification with resnext. In: IEEE
  Congress on Evolutionary Computation. pp.~1--7 (2020)

\bibitem{8887303}
He, K., Kim, D.S.: Malware detection with malware images using deep learning
  techniques. In: 18th IEEE International Conference On Trust, Security And
  Privacy In Computing And Communications/13th IEEE International Conference On
  Big Data Science And Engineering. pp. 95--102 (2019)

\bibitem{densenet}
Huang, G., Liu, Z., Maaten, L.V.D., Weinberger, K.Q.: Densely connected
  convolutional networks. In: 2017 IEEE Conference on Computer Vision and
  Pattern Recognition (CVPR). pp. 2261--2269. IEEE Computer Society, Los
  Alamitos, CA, USA (jul 2017). \doi{10.1109/CVPR.2017.243}

\bibitem{IADAROLA2021102198}
Iadarola, G., Martinelli, F., Mercaldo, F., Santone, A.: Towards an
  interpretable deep learning model for mobile malware detection and family
  identification. Computers \& Security  \textbf{105},  102198 (2021)

\bibitem{JERBI2020101743}
Jerbi, M., Dagdia, Z.C., Bechikh, S., Said, L.B.: On the use of artificial
  malicious patterns for android malware detection. Computers \& Security
  \textbf{92},  101743 (2020)

\bibitem{JIAN2021102400}
Jian, Y., Kuang, H., Ren, C., Ma, Z., Wang, H.: A novel framework for
  image-based malware detection with a deep neural network. Computers \&
  Security  \textbf{109},  102400 (2021)

\bibitem{9356071}
Jin, X., Xing, X., Elahi, H., Wang, G., Jiang, H.: A malware detection approach
  using malware images and autoencoders. In: IEEE 17th International Conference
  on Mobile Ad Hoc and Sensor Systems. pp.~1--6 (2020)

\bibitem{motif}
Joyce, R.J., Amlani, D., Nicholas, C., Raff, E.: Motif: A malware reference
  dataset with ground truth family labels. Computers \& Security  \textbf{124},
   102921 (2023)

\bibitem{Jung:2018}
Jung, J., Kim, H., Shin, D., Lee, M., Lee, H., Cho, S.j., Suh, K.: Android
  malware detection based on useful api calls and machine learning. In: IEEE
  1st International Conference on Artificial Intelligence and Knowledge
  Engineering. pp. 175--178 (2018)

\bibitem{KABAKUS2022117833}
Kabakus, A.T.: Droidmalwaredetector: A novel android malware detection
  framework based on convolutional neural network. Expert Systems with
  Applications  \textbf{206},  117833 (2022)

\bibitem{10.1007/978-3-030-80825-9_16}
Karbab, E.B., Debbabi, M.: Petadroid: Adaptive android malware detection using
  deep learning. In: Bilge, L., Cavallaro, L., Pellegrino, G., Neves, N. (eds.)
  Detection of Intrusions and Malware, and Vulnerability Assessment. pp.
  319--340 (2021)

\bibitem{MAPAS}
Kim, J., Ban, Y., Ko, E., Cho, H., Yi, J.H.: Mapas: a practical deep
  learning-based android malware detection system. International Journal of
  Information Security  \textbf{21},  725–738 (2022)

\bibitem{8443370}
Kim, T., Kang, B., Rho, M., Sezer, S., Im, E.G.: A multimodal deep learning
  method for android malware detection using various features. IEEE Trans. on
  Information Forensics and Security  \textbf{14}(3),  773--788 (2019)

\bibitem{KONG2022102514}
Kong, K., Zhang, Z., Yang, Z.Y., Zhang, Z.: Fcscnn: Feature centralized siamese
  cnn-based android malware identification. Computers \& Security
  \textbf{112},  102514 (2022)

\bibitem{KUMAR2022103063}
Kumar, S., Janet, B.: Dtmic: Deep transfer learning for malware image
  classification. Journal of Information Security and Applications
  \textbf{64},  103063 (2022)

\bibitem{cic_mal_2017}
Lashkari, A.H., Kadir, A.F.A., Taheri, L., Ghorbani, A.A.: Toward developing a
  systematic approach to generate benchmark android malware datasets and
  classification. In: International Carnahan Conference on Security Technology.
  pp.~1--7 (2018)

\bibitem{Li:2018}
Li, J., Sun, L., Yan, Q., Li, Z., Srisa-an, W., Ye, H.: Significant permission
  identification for machine-learning-based android malware detection. IEEE
  Trans. on Industrial Informatics  \textbf{14}(7),  3216--3225 (2018)

\bibitem{LIU2020101682}
Liu, X., Lin, Y., Li, H., Zhang, J.: A novel method for malware detection on
  ml-based visualization technique. Computers \& Security  \textbf{89},  101682
  (2020)

\bibitem{8629067}
Ma, Z., Ge, H., Liu, Y., Zhao, M., Ma, J.: A combination method for android
  malware detection based on control flow graphs and machine learning
  algorithms. IEEE Access  \textbf{7},  21235--21245 (2019).
  \doi{10.1109/ACCESS.2019.2896003}

\bibitem{Mahindru:2021}
Mahindru, A., Sangal, A.L.: Mldroid---framework for android malware detection
  using machine learning techniques. Neural Computing and Applications
  \textbf{33}(10),  5183--5240 (May 2021)

\bibitem{Mercaldo:2020}
Mercaldo, F., Santone, A.: Deep learning for image-based mobile malware
  detection. Journal of Computer Virology and Hacking Techniques
  \textbf{16}(2),  157--171 (Jun 2020)

\bibitem{3c2d}
Mohammed, T.M., Nataraj, L., Chikkagoudar, S., Chandrasekaran, S., Manjunath,
  B.: Malware detection using frequency domain-based image visualization and
  deep learning. In: 54th Hawaii International Conference on System Sciences.
  p.~7132 (2021)

\bibitem{malimg}
Nataraj, L., Karthikeyan, S., Jacob, G., Manjunath, B.S.: Malware images:
  Visualization and automatic classification. In: 8th International Symposium
  on Visualization for Cyber Security (2011)

\bibitem{NISSIM20145843}
Nissim, N., Moskovitch, R., Rokach, L., Elovici, Y.: Novel active learning
  methods for enhanced pc malware detection in windows os. Expert Systems with
  Applications  \textbf{41}(13),  5843--5857 (2014)

\bibitem{10.1145/3313391}
Onwuzurike, L., Mariconti, E., Andriotis, P., Cristofaro, E.D., Ross, G.,
  Stringhini, G.: Mamadroid: Detecting android malware by building markov
  chains of behavioral models (extended version). ACM Trans. Priv. Secur.
  \textbf{22}(2) (2019)

\bibitem{OU2022102513}
Ou, F., Xu, J.: S3feature: A static sensitive subgraph-based feature for
  android malware detection. Computers \& Security  \textbf{112},  102513
  (2022)

\bibitem{PEI2020101792}
Pei, X., Yu, L., Tian, S.: Amalnet: A deep learning framework based on graph
  convolutional networks for malware detection. Computers \& Security
  \textbf{93},  101792 (2020)

\bibitem{Pekta:2020}
Pekta, A., Acarman, T.: Deep learning for effective android malware detection
  using api call graph embeddings. Soft Computing  \textbf{24}(2),  1027--1043
  (2020)

\bibitem{PINHERO2021102247}
Pinhero, A., {M L}, A., P, V., Visaggio, C., N, A., S, A., S, A.: Malware
  detection employed by visualization and deep neural network. Computers \&
  Security  \textbf{105},  102247 (2021)

\bibitem{malconv}
Raff, E., Fleshman, W., Zak, R., Anderson, H.S., Filar, B., McLean, M.:
  Classifying sequences of extreme length with constant memory applied to
  malware detection. In: AAAI Conference on Artificial Intelligence. pp.
  9386--9394 (2021)

\bibitem{10.1145/3442520.3442522}
Rahali, A., Lashkari, A.H., Kaur, G., Taheri, L., GAGNON, F., Massicotte, F.:
  Didroid: Android malware classification and characterization using deep image
  learning. In: 10th International Conference on Communication and Network
  Security. p. 70–82 (2021)

\bibitem{ronen2018microsoft}
Ronen, R., Radu, M., Feuerstein, C., Yom-Tov, E., Ahmadi, M.: Microsoft malware
  classification challenge (2018)

\bibitem{mobilenetv2}
Sandler, M., Howard, A., Zhu, M., Zhmoginov, A., Chen, L.C.: Mobilenetv2:
  Inverted residuals and linear bottlenecks (2019)

\bibitem{SASIDHARAN2021101336}
Sasidharan, S.K., Thomas, C.: Prodroid — an android malware detection
  framework based on profile hidden markov model. Pervasive and Mobile
  Computing  \textbf{72},  101336 (2021)

\bibitem{VGG16}
Simonyan, K., Zisserman, A.: Very deep convolutional networks for large-scale
  image recognition (2014)

\bibitem{SURENDRAN2020102483}
Surendran, R., Thomas, T., Emmanuel, S.: A tan based hybrid model for android
  malware detection. Journal of Information Security and Applications
  \textbf{54},  102483 (2020)

\bibitem{TEKEREK2022102515}
Tekerek, A., Yapici, M.M.: A novel malware classification and augmentation
  model based on convolutional neural network. Computers \& Security
  \textbf{112},  102515 (2022)

\bibitem{VASAN2020107138}
Vasan, D., Alazab, M., Wassan, S., Naeem, H., Safaei, B., Zheng, Q.: Imcfn:
  Image-based malware classification using fine-tuned convolutional neural
  network architecture. Computer Networks  \textbf{171},  107138 (2020)

\bibitem{VASAN2020101748}
Vasan, D., Alazab, M., Wassan, S., Safaei, B., Zheng, Q.: Image-based malware
  classification using ensemble of cnn architectures ({IMCEC}). Computers \&
  Security  \textbf{92},  101748 (2020)

\bibitem{HIT4Mal}
Vu, D.L., Nguyen, T.K., Nguyen, T.V., Nguyen, T.N., Massacci, F., Phung, P.H.:
  Hit4mal: Hybrid image transformation for malware classification. Trans. on
  Emerging Telecommunications Technologies pp. 1 -- 15 (2019)

\bibitem{9631209}
Wong, W.K., Juwono, F.H., Apriono, C.: Vision-based malware detection: A
  transfer learning approach using optimal ecoc-svm configuration. IEEE Access
  \textbf{9},  159262--159270 (2021)

\bibitem{XIAO2021102420}
Xiao, M., Guo, C., Shen, G., Cui, Y., Jiang, C.: Image-based malware
  classification using section distribution information. Computers \& Security
  \textbf{110},  102420 (2021)

\bibitem{10.1007/978-3-030-02450-5_11}
Xu, Z., Ren, K., Qin, S., Craciun, F.: Cdgdroid: Android malware detection
  based on deep learning using cfg and dfg. In: Sun, J., Sun, M. (eds.) Formal
  Methods and Software Engineering (2018)

\bibitem{YADAV2022102622}
Yadav, P., Menon, N., Ravi, V., Vishvanathan, S., Pham, T.D.: Efficientnet
  convolutional neural networks-based android malware detection. Computers \&
  Security  \textbf{115},  102622 (2022)

\bibitem{YUAN2020101740}
Yuan, B., Wang, J., Liu, D., Guo, W., Wu, P., Bao, X.: Byte-level malware
  classification based on markov images and deep learning. Computers \&
  Security  \textbf{92},  101740 (2020)

\bibitem{shufflenet}
Zhang, X., Zhou, X., Lin, M., Sun, J.: Shufflenet: An extremely efficient
  convolutional neural network for mobile devices. In: 2018 IEEE/CVF Conference
  on Computer Vision and Pattern Recognition (CVPR). pp. 6848--6856. IEEE
  Computer Society, Los Alamitos, CA, USA (jun 2018)

\bibitem{10.1145/3372297.3417291}
Zhang, X., Zhang, Y., Zhong, M., Ding, D., Cao, Y., Zhang, Y., Zhang, M., Yang,
  M.: Enhancing state-of-the-art classifiers with api semantics to detect
  evolved android malware. In: ACM SIGSAC Conference on Computer and
  Communications Security. p. 757–770 (2020)

\bibitem{nver:2020}
Ünver, H.M., Bakour, K.: Android malware detection based on image-based
  features and machine learning techniques. SN Applied Sciences  \textbf{2}(7),
  ~1299 (2020)

\end{thebibliography}

\end{document}